\documentclass[journal, onecolumn]{IEEEtran}
\usepackage{cite}
\usepackage{amsmath,amssymb,amsfonts}
\usepackage{algorithmic}
\usepackage{graphicx}
\usepackage{textcomp}

\usepackage{algorithm,algorithmic}
\usepackage{float}
\usepackage{color,soul}

 \ifCLASSINFOpdf
  
\else

\fi

\begin{document} 

\hyphenation{op-tical net-works semi-conduc-tor}

\title{Time Reversal based MAC for Multi-Hop Underwater Acoustic Networks}

\author{Ruiqin~Zhao,~Hao~Long,~
				Octavia~A.~Dobre,
				Xiaohong~Shen,~
				Telex~M.~N.~Ngatched,~
				and~Haodi~Mei
        
\thanks{Ruiqin Zhao,Hao Long,Xiaohong Shen andHaodi Mei are with the Key Laboratory of Ocean Acoustics and Sensing, School of Marine Science and Technology, Northwestern Polytechnical University, Xi’an 710072, China (e-mail:, rqzhao@nwpu.edu.cn).}
\thanks{Octavia A. Dobre and Telex M. N. Ngatched are with Faculty of Engineering and Applied Science, Memorial University, St. John’s, NL A1B 3X5, Canada (e-mail:,odobre@mun.ca; tngatched@grenfell.mun.ca). }
\thanks{This work was supported in part by the National Natural Science Foundation of China under Grants 61571367 and 61671386, in part by the National Key Program of China under Grant 2016YFC1400200, and in part by the Memorial University Chair on Subsea Communications.}

}

%



\maketitle

\begin{abstract}
Constrained-energy underwater acoustic nodes are typically connected via a multi-hop underwater acoustic network (MHUAN) to cover a broad marine region. Recently, protocols for efficiently connecting such nodes have received considerable attention. In this paper, we show that the time reversal (TR) process plays an important role in the medium access control (MAC) because of its physical capability to exploit the multi-path energy from the richly scattering underwater environment, as well as to focus the signal energy in both spatial and temporal domains. In MHUANs, with severe multi-path propagation at the physical layer, the active TR process spatially focuses the signals to the location of the intended receiver; this significantly diminishes the interference among parallel links. We propose an active TR-based MAC protocol for MHUANs, with the aim of minimizing collision and maximizing channel utilization simultaneously. Furthermore, by considering the impact of the cross-correlation between different links on the TR-based medium access, we derive the threshold of the link cross-correlation to resolve collision caused by the high cross-correlation between realistic links. We perform simulations using the OPNET and BELLHOP environments, and show that the proposed TR-based MAC results in significantly improved throughput, decreased delay, and reduced data drop ratio in MHUANs.
\end{abstract}



%
\IEEEpeerreviewmaketitle

\section{Introduction}
The last decade has witnessed rapidly growing interest in underwater acoustic networks, owing to their broad applications, such as ocean sampling, environmental monitoring, undersea exploration, disaster prevention, assisted navigation, and distributed tactical surveillance. These networks consist of sensors and vehicles deployed underwater and connected via acoustic links to perform collaborative tasks. The deployment of the network nodes can be over a large area, and by considering the energy constraint in such networks, multi-hop communications are required. These networks are referred to as multi-hop underwater acoustic networks (MHUANs).

\subsection{Motivation}
Due to the harsh marine environment and unique features of the underwater acoustic channel at the physical layer, MHUANs have particular characteristics when compared with terrestrial wireless multi-hop networks, as follows: 1) Due to the severe multi-path effect of underwater acoustic channel, the link data rate is significantly reduced \cite{4752682,4445738,7404366,7127542}; 2)  As the nominal propagation speed of the acoustic signal is 1500 m/s, the propagation delay in MHUANs is large, and significantly affects the network performance, such as throughput, delay, link utilization efficiency and energy consumption; 3) Owing to the complex propagation environment, the transmit power of the underwater acoustic modems is orders of magnitude higher than that of the terrestrial wireless node. For example, the typical transmit power for the Woods Hole Oceanographic Institution (WHOI) micro-modem is no less than 30 W \cite{WHOI}, while a radio frequency sensor has a transmit power around 80 mW \cite{Kredo:2007:HMA:1287812.1287821}. If a transmitted data frame is not received successfully, the same data frame is retransmitted to guarantee correct reception. Thus, when considering the limited energy at each node, a high successful transmission ratio needs to be ensured in MHUANs; this is a key factor for high energy efficiency, long lifetime, short end-to-end delay, and high throughput.   

In order to improve the success of the transmission, different strategies at the physical, medium access control (MAC), and logical link control (LLC) layers have been considered. At the MAC layer, protocols coordinate the node access to the shared broadcast channel; therefore, with the aim of achieving a high successful transmission ratio, collisions need to be minimized. Existing MAC protocols for MHUANs can be mainly divided into two classes: i)  contention-based protocols, such as carrier sense multiple access with collision avoidance (CSMA/CA) \cite{820738}, CSMA/CA-based \cite{6086556,ShumingXiong2013RANF}, and ALOHA-based; ii) contention-free protocols, such as frequency division multiple access, time division multiple access (TDMA), and code division multiple access (CDMA). The former accesses channel resources either through handshake-type mechanisms with control messages, or by exploiting the access randomness. Due to their random access feature, collisions can not be completely avoided when contention-based protocols are applied. On the other hand, the latter allocates channel resources in a predefined way and minimizes the collision at the cost of extra constraints. For example, time synchronization is needed in TDMA, while alleviating the interference caused by the near-far effect is required to ensure conflict-free sharing of the common medium in CDMA.
 
The harsh underwater environment causes complex and unique spatially varying features of the underwater acoustic channel\cite{4752682,7299704,7346509}, resulting in severe multi-path effect, and thus, very limited data rate. However, these features can be utilized by the time reversal (TR) process \cite{7299704,6964573}, which focuses the signal energy in both spatial and temporal domains. Therefore, researchers have used TR as a method to increase the communication data rate or to multiplex several users \cite{7926399,4533497,7996470}. In terrestrial wireless networks, a multiuser downlink communication strategy based on TR multiplexing was presented as a solution to energy-efficient broadband wireless communication in \cite{7398145}. Laboratory TR experiments have shown the spatial focusing and temporal compression across a broad range of settings \cite{4655408,4098939,4445738}. In \cite{4098939}, the authors have utilized the TR process to explore the spatial focusing and temporal compression potential in underwater acoustic multiple-input and multiple-output (MIMO) communication. They established active and passive TR-based communication systems for the downlink and uplink, respectively. The temporal compression of the TR process mitigates the inter-symbol interference (ISI), while its spatial focusing enables the extension to multiuser MIMO communications \cite{4445738}. TR-based transmission plays an important role in low-complexity energy-efficient communications, exploiting the multi-path energy from the richly scattering underwater environment and focusing the signal energy at the intended location \cite{5992838,7891649}. However, all existing works \cite{7398145,4533497,6964573,4098939,5992838,7891649,4445738,7926399,7996470} using TR in underwater networks have either been done at the physical layer to improve data rate or in a one-hop network scenario.

\subsection{Contributions}
We investigate the MAC strategy for MHUANs utilizing the TR process in this paper, with the aim of minimizing the collision and improving the channel utilization at the same time. Our contributions are as follows:

1) The impacts of the active and passive TR processes on MAC are compared and evaluated in the multi-hop network scenario. The active TR process is then selected since its spatial focusing capability isolates the interference during simultaneous transmissions.

2) For the multi-hop network scenario, a TR-based MAC protocol (TRMAC) is proposed. Exploiting the TR nature of harvesting the multi-path energy from the richly scattering underwater environment and concentrating the signal energy at the intended node in the network, the TRMAC grants parallel transmissions with significantly diminished interference.

3) Inspired by the similarity of the probe reservation required by the active TR process and the channel reservation of the MAC strategy, we combine them in a smart way, enabling random access based on the active TR probe reservation and resulting in an improved channel utilization in MHUANs.

4) To remove the residual collision caused by the potentially high cross-correlation between realistic links, the threshold of the link cross-correlation is derived by evaluating its impact on the received signal-to-interference-plus-noise ratio (SINR) when parallel TR links simultaneously access the shared medium.

The rest of this paper is organized as follows: Section II discusses the application of the TR process to the MAC mechanism for MHUANs. The system model is introduced in Section III, and the proposed TRMAC protocol is presented in Section IV, including the cross-layer design for TRMAC, the algorithm description, as well as the derivation of the threshold of the cross-correlation between links. Section V presents the performance evaluation of the proposed protocol by using the OPNET and BELLHOP simulation environments, and Section VI concludes the paper. 

\vspace{0.3cm}

\section{Applying TR Process to MHUAN MAC}
\subsection{Active TR vs. Passive TR for MAC}
The TR process can be applied to underwater acoustic networks in two manners, namely passive and active TR. 

For the passive TR, the transmitter node sends the piloting probe and data signals, which are separated by a short interval. After receiving them, the receiver completes the TR process for data reception based on the piloting probe signal that has recorded the diversity of channel paths. In this way, any node within the communication range of the transmitter node can obtain the signal-to-noise ratio (SNR) gain from the TR process through the received piloting probe signal. 

For the active TR, the transmitter node firstly gets the record of the channel paths diversity through a pilot signal requested from its receiver node. Then, based on this record of the channel spatial diversity, it completes the TR process that uses the time-reversed channel impulse response (CIR) as the basic waveform to transmit data signals through the same channel to the receiver node; this is equivalent to setting a private mark to the data signals on this link. Due to channel reciprocity, the time-reversed signals can retrace the incoming paths, ending up with a resulting "spiky" spatial signal-power distribution focused at the intended location \cite{4655408,5992838}. Therefore, within the communication range of the transmitter node, due to the spatially varying channels in MHUANs, only the intended receiver can obtain the SNR gain from the active TR process.

The active and passive TR processes are theoretically equivalent when the SNR of the intended receiver is evaluated at the physical layer. However, when they are investigated at the MAC layer, the difference is significant. When the active TR is used, the "spiky" signal only focuses at the intended receiver node, and little interference is formed on other nodes in the network\cite{5992838}. 
This spatial focusing capability of the active TR isolates the interference between adjacent links, and grants their parallel transmissions with collision avoidance in underwater acoustic networks. Thus, this type of process is used  at the physical layer in this paper to exploit its spatial focusing capability in MHUANs.

Furthermore, the TR focusing performance can be improved by applying an array at the transceiver \cite{1707996,7398145}. Considering the node cost and size, in this paper, we discuss the TR transmission with a single transducer element.
\subsection{Impact of Link Cross-correlation on MAC}
In order to explore and observe over a broad region, MHUAN nodes are deployed in a sparse pattern. Generally, nodes are deployed with diverse depths and spaced by various distances. This leads to low cross-correlation between different links \cite {4655408,7398145}, and yields the interference isolation between adjacent links.

However, due to the spatial similarity between links and some environmental factors, the cross-correlation between adjacent links may become strong. In such cases, the collision appears among the TR-based simultaneous transmissions. The existing works typically do not consider the impact of the cross-correlation between different links. A solution to this problem is provided in Section IV. 
\section{System Model}
An MHUAN can be abstracted as a graph $G(V,E)$, in which $V$ is the set of nodes in the network and $E$ consists  of the links in the network. Each node is in half-duplex mode and can only receive signals from one node at a time.

\begin{figure}[htbp]
 \centering
 \includegraphics[width=3.5 in]{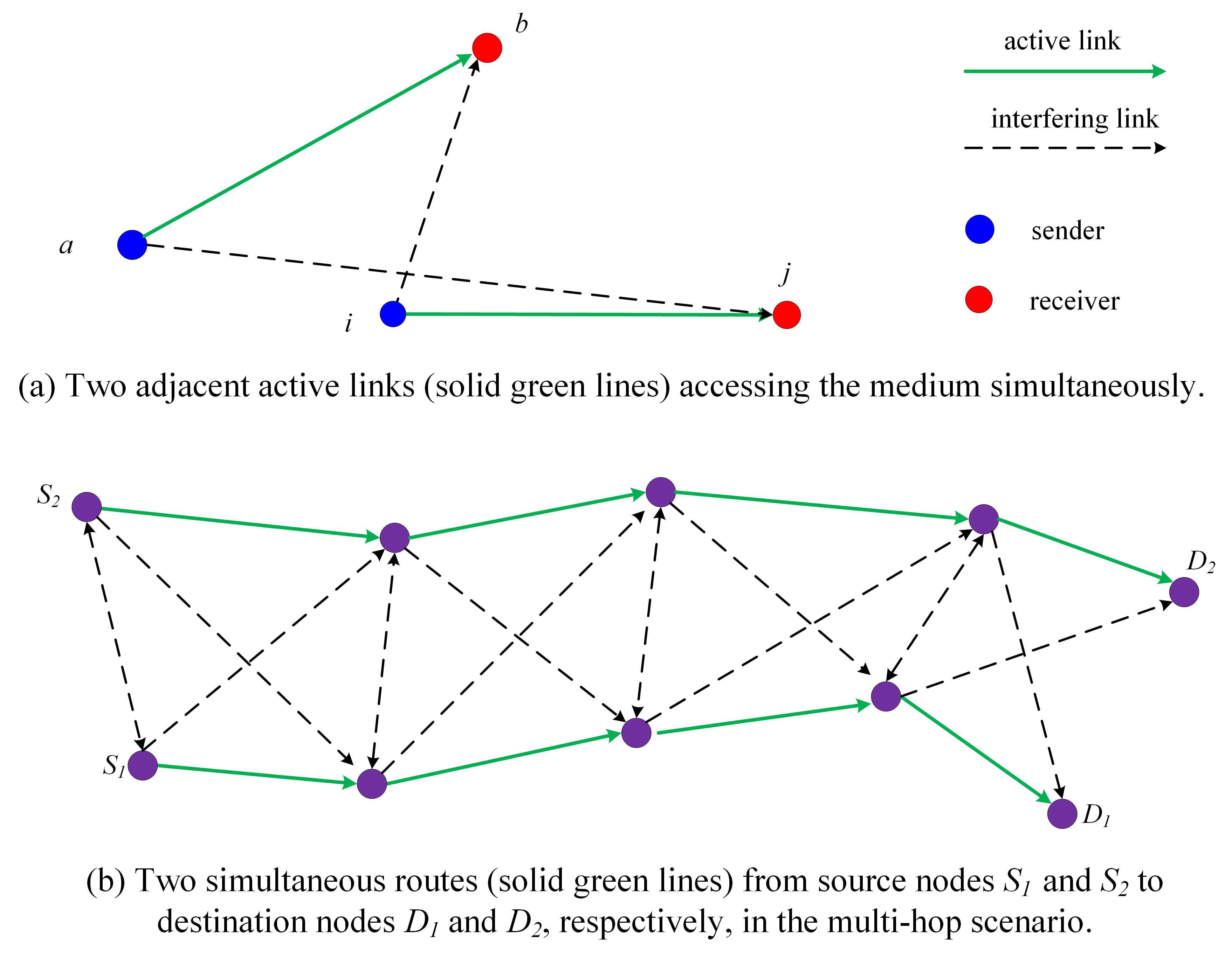}
 \caption{Parallel active links (solid green lines) accessing the medium simultaneously in MHUAN.}
\label{fig1}
\end{figure}

The network model is presented in Fig. 1, where active links are shown by solid green lines and the induced interfering links are displayed by black dashed lines. In Fig. 1 (a), two adjacent active links from node $a$ to node $b$ and from node $i$ to node $j$ share the channel simultaneously. Because the data signal of each active link is time-reversed based on its link CIR before being transmitted, the signal energy received at the intended node is focused in both spatial and temporal domains; this enables simultaneous active links. Thus, in the multi-hop scenario shown in Fig. 1 (b), multiple parallel routes from different source nodes to different destination nodes can be established without inter-route interference. Therefore, the number of simultaneous active links that do not interfere with each other significantly increases, and the network throughput significantly improves.

At the physical layer, the CIR at time $k$ of the link between node $a$ and node $b$ in discrete time domain is modeled as
\begin{equation}h_{ab}[k]=\sum_{l=0}^{L-1}h_{l}^{ab}\cdot \delta [k-l],  \end{equation}
where $h_{l}^{ab}$ is the complex amplitude of the $l$-th tap of the CIR for link $ab$, $L$ is the number of channel taps, and $\delta[.]$ is the Dirac delta function. Assuming slow-varying fading, the channel taps are constant during the observation time; the CIR of link $ab$ can be represented as
\begin{equation}\mathbf{h}^{ab}=\left[h_{0}^{ab}\; h_{1}^{ab}\; \cdot\cdot\cdot \; h_{L-1}^{ab}\right]^{\dag},\end{equation}
where the superscript $\dag$ represents the transpose operator. 

In $G(V,E)$, the cross-correlation between the CIRs of the two links $ij$ and $pq$ ($\forall ij, pq \in E$) is defined as\cite{proakisdsp}
\begin{equation}r_{ij,pq}[k]=\sum_{l=0}^{L-1}h_{ij}[l]\cdot h_{pq}^{*}[l+k],   k=0, \pm1, \pm2.... \end{equation} In order to let the cross-correlation be no larger than 1, the normalized cross-correlation is defined as
\begin{equation}\eta_{ij,pq}[k]=\frac{r_{ij,pq}[k]}{\left \| \mathbf{h}^{ij} \right \|\cdot \left \|\mathbf{h}^{pq}\right \|},\end{equation} 
where $\left \| \mathbf{h}^{ij} \right \|$ and $\left \| \mathbf{h}^{pq} \right \|$ are the norms of $\mathbf{h}^{ij}$ and $\mathbf{h}^{pq}$, respectively. 

Based on (3) and (4), one can infer that the normalized cross-correlation between different links is an important factor that describes the spatial correlation between links in networks. The proposed TRMAC uses the cross-correlation between links to efficiently coordinate channel access of nodes with the aim of collision avoidance in MHUANs.

\begin{figure*}[h]
 \centering
 \includegraphics{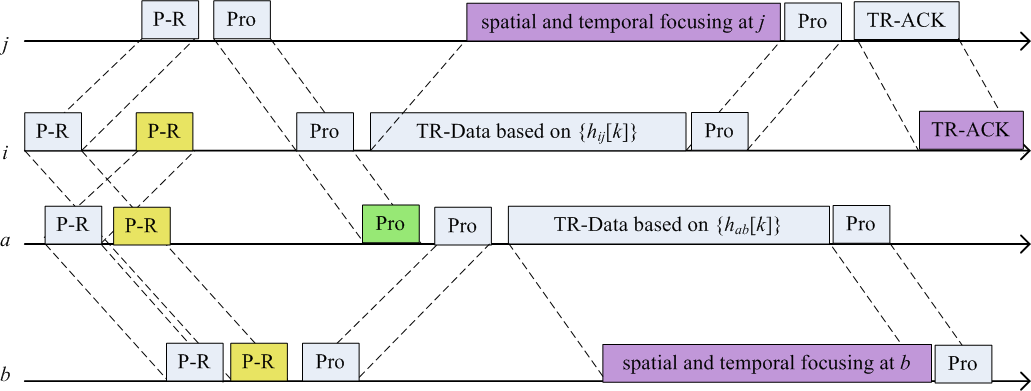}
 \caption{Channel reservation and recording through P-R/Pro/TR-Data with Pro/TR-ACK handshake in TRMAC.}
\label{fig2}
\end{figure*}
  
\section{Proposed TRMAC Protocol}
A TRMAC strategy is proposed for MHUANs with the aim of minimizing collision and improving channel utilization. TRMAC enables random access based on the active TR probe reservation and results in an improved channel utilization in MHUANs. To resolve the residual collision caused by the potentially high cross-correlation between realistic links, the threshold of the link cross-correlation is derived by evaluating its impact on the SINR when parallel TR links simultaneously access the medium. If the threshold is exceeded, one transmission is delayed in the TRMAC. Based on the cross-layer design for TRMAC and the derived threshold of cross-correlation between links, the proposed TRMAC strategy can minimize the collision and improve channel utilization. 

The proposed TRMAC uses four frames, which are defined as follows:
\begin{itemize}
	\item P-R: Probe Request frame.
	\item Pro: Probe frame used for active TR.
	\item TR-Data: Data frame Time-Reversed before being transmitted. 
	\item TR-ACK: Acknowledgment (ACK) frame Time-Reversed before being transmitted.
\end{itemize}

\subsection{Cross-layer Design in TRMAC}
Generally, the active TR transmission consists of two stages: the channel recording stage and data transmission stage. In the first stage, the transmitter obtains the record of diversity of the channel paths through a probe signal from its receiver. In the second stage, the transmitter simply uses the time-reversed CIR as a basic waveform to transmit data signals through the same channel to the receiver. 

As the active TR is adopted at the physical layer, the probe reservation for channel recording is required, similar to the channel reservation process in MAC protocols. Inspired by this, we accomplish both channel recording required by the active TR process at the physical layer and channel reservation needed at the MAC layer through the handshake of P-R and Pro. In the TRMAC, as shown in Fig. 2, the random medium access is achieved based on the handshake of P-R/Pro/TR-Data with Pro/TR-ACK. Similar to the active TR process, the TRMAC consists of two steps: channel reservation and recording step, and TR-based link transmission step. 

\textit{\textbf{Step 1: Channel Reservation and Recording}} 

As shown in Fig. 2, $\forall a\in V$, when node $a$ needs to transmit data to node $b$, it sends a P-R frame including a probe signal to node $b$ to acquire the probe signal required by the active TR process and to reserve the channel. After receiving the P-R frame, node $b$ can obtain the CIR of the link $ab$ using the included probe signal, which could be used to calculate the link cross-correlation threshold. 

If not reserved by other links, node $b$ would reply with a Pro frame which propagates through the underwater acoustic channels featured by severe multi-path effect and spatial variability. After receiving the Pro frame, node $a$ not only successfully reserves the channel, but also calculates the CIR of link $ab$ based on the recorded waveform of the received Pro. Then, the transmitter node $a$ gets
\begin{equation}g_{ab}[k]=\frac{h_{ab}^{*}[L-1-k]}{\left \| \mathbf{h}^{ab} \right \|},\end{equation}
where $g_{ab}[k]$ ($k = 0, 1, 2,...,L-1$) represents the normalized time-reversed CIR required by the TR-based link transmission step for link $ab$. For simplicity of analytical derivation, we assume that $h_{ab}[k]$ in (5) represents the true value of the CIR of link $ab$. 

\textit{\textbf{Step 2: TR-based Link Transmission}} 

The transmitter node $a$ uses the normalized time-reversed CIR of link $ab$, $\left\{g_{ab}[k]\right\}$, as the waveform to transmit the data frame through the same channel to its intended receiver node $b$. In order to reduce the residual ISI at the receiver, up-sampling and down-sampling are introduced at the transmitter and receiver, respectively. Let $\left\{x_{a}[k]\right\}$ be the data symbol sequence generated from node $a$, which is firstly up-sampled at the transmitter as\cite{5992838} 
\begin{equation}x_{a}^{D}[k]=\begin{cases}
 & x_{a}[k/D], \quad\text{ if } k \; mod\; D = 0, \\ 
 & \quad\; 0,\quad\quad\;\text{ if }  k\; mod\; D \neq  0,
\end{cases}\end{equation}
where $D$ is the up/down-sampling factor.\footnote {The introduction of $D$ would result in the increase of the required bandwidth when $D > 1$. Under the assumption of the same data rate, it requires $D$ times the bandwidth when compared with $D = 1$. The impact of $D$ will be discussed in detail in Section IV. A.} 
The up-sampled data sequence is time-reversed through its convolution with the TR waveform $\left\{g_{ab}[k]\right\}$. Thus, the signal arriving at the intended node can be expressed as  
\begin{equation}
y^{D}[k]=x_{a}^{D}[k]\ast g_{ab}[k]\ast h_{ab}[k]+n[k],
\end{equation}
where $\left\{n[k]\right\}$ is the sequence of the complex-valued additive white Gaussian noise samples with zero-mean and variance $\sigma^{2}$. 

Accordingly, the receiver down-samples the arriving signal with the same factor $D$. The down-sampled signal is expressed as
\begin{equation}
y[k]=\sum_{l=0}^{2(L-1)/D}(h_{ab}*g_{ab})[Dl] x_{a}[k-l]+n[Dk],
\end{equation} 
where $\left\{n[Dk]\right\}$ is the sequence of the complex-valued additive white Gaussian noise samples with zero-mean and variance $\sigma^{2}$, $L-1$ is assumed to be dividable by $D$ for simplicity, and
\begin{equation}(h_{ab}*g_{ab})[k]=\frac{h_{ab}[k]\ast h_{ab}^{*}[L-1-k]}{\left \| \mathbf{h}^{ab} \right \|}.\end{equation}

Based on the definition of convolution and using (3) and (4), $\forall ij, pq\in E$, one can obtain the relation between the active TR process and the cross-correlation between links as
\begin{equation}
\begin{split}
h_{ij}[k]\ast h_{pq}^{*}[L-1-k]&=r_{ij,pq}[k-(L-1)]\\
&=\eta_{ij,pq}[k-(L-1)] \left \| \mathbf{h}^{ij} \right \| \left \|\mathbf{h}^{pq}\right \|.
\end{split}
\end{equation}
Thus, with (8), (9), and (10), the received signal at the intended node is given by
\begin{equation}\begin{split}
y[k]&=\left \|\mathbf{h}^{ab}\right \|\sum_{l=0}^{(2L-1)/D}\eta_{ab,ab}[Dl-(L-1)] x_{a}[k-l]+n[Dk] \\
&=\left \|\mathbf{h}^{ab}\right \| \eta_{ab,ab}[0] x_{a}[k] \quad \quad\quad\quad\quad\quad\quad\mbox{(Signal)}\\
&+\left \|\mathbf{h}^{ab}\right \| \sum_{l= 0; \; l\neq (L-1)/D}^{2(L-1)/D}\eta_{ab,ab}[Dl-(L-1)] x_{a}[k-l] \mbox{(ISI)}\\
&+n[Dk].\quad\quad\quad\quad\quad\quad\quad\quad\quad\quad\quad\quad\quad\;\mbox{(Noise)}
\end{split}\end{equation} 
As shown in (4), $\eta_{ab,ab}[k]$ is the auto-correlation function of the CIR of the link between nodes $a$ and $b$. When $l=(L-1)/D$, $\eta_{ab,ab}[Dl-(L-1)]$ equals $\eta_{ab,ab}[0]$, achieving its maximum value of 1.

From (11), one can calculate the signal power, $P_{\texttt{Sig}}$, and the ISI power, $P_{\texttt{ISI}}$, as
\begin{equation}P_{\texttt{Sig}}=DP \left \|\mathbf{h}^{ab}\right \|^{2},\end{equation}
and
\begin{equation}P_{\texttt{ISI}}=DP \left \|\mathbf{h}^{ab}\right \|^{2}\sum_{l= 0; \; l\neq (L-1)/D}^{2(L-1)/D}\left|\eta_{ab,ab}[Dl-(L-1)]\right|^{2},\end{equation}
where $P$ is the average transmit power of $\left\{x_{a}^{D}[k]\right\}$. In the same way, the interference at node $b$ caused by the transmission on link $ij$ in Fig. 1 can be expressed as
\begin{equation}y^{'}[k]=\left \|\mathbf{h}^{ib}\right \|\sum_{l=0}^{2(L-1)/D}\eta_{ib,ij}[Dl-(L-1)] x_{i}[k-l]+n[Dk],\end{equation} 
where $\left\{x_{i}[k]\right\}$ is the data symbol sequence from node $i$. This interference is called inter-link interference (ILI). As shown in (11) and (14), both the received signal and interference are determined by the normalized cross-correlation between links; thus, the TRMAC uses it to evaluate whether a collision will be induced during the parallel transmissions, with the aim of collision avoidance.

Due to the typically low correlation between links in the network, caused by the unique spatial variety feature of the underwater acoustic channel, the cross-correlation between different links is generally low, resulting in a reduced interference. Hence, the up/down-sampling method leads to interference reduction with the factor $D > 1$. Thus, the transmitted TR-Data signal focuses on the position of the intended node, which also shows the spatial focusing of the active TR-based transmission.

During the P-R/Pro/TR-Data with Pro/TR-ACK handshake shown in Fig. 2, a new Pro following the TR-Data is used to record the updated CIR of the link. When this Pro arrives at the receiver, the updated link CIR is exploited to transmit the ACK frame in a TR-based way to improve the transmission reliability. The multi-path channel forms a natural matched filter to the basic waveform of the time-reversed Data and ACK, thus enabling multiple links to simultaneously access the common channel without interfering with each other. 

Furthermore, in order to resolve the residual collision caused by the potentially high cross-correlation between realistic links, the threshold of the cross-correlation between links is derived in Section III C. In the following subsection, a detailed description of the proposed MAC protocol is presented. 

\subsection{Algorithm Description}

The variables used in the algorithm are defined as follows:

$t_{p}$: maximum propagation delay of each hop in the network.

$t_{tr}$: transmission delay of the TR-Data frame, which equals the TR-Data frame length divided by the data rate.

$\Delta$: guard time, which includes the transmission delay of the reply frames, like Pro and TR-ACK, from the receiver.

$T_{cl}$: interference interval or collision interval for an on-going transmission; $T_{cl}=t_{p}+t_{tr}+\Delta$.

$T_{th}$: threshold of a waiting interval for retransmission; nodes would retransmit the P-R or TR-Data frame if they have not received the reply from the receiver node after the waiting interval exceeds $T_{th}$; $T_{th}=2t_{p}+t_{tr}+\Delta$.

$T$: coherence time of the channel, which can be set based on the prior measurements of the realistic marine environment of the network deployment.

$T_{Pro}^{j}$: interval from the reception of the overheard Pro frame generated by node $j$ to current time.

$N_{max}$: maximum number of retransmissions during medium access. If the number of retransmissions for a certain packet exceeds the limit $N_{max}$, the sender node drops the packet.

$\eta_{th}$: threshold for the normalized cross-correlation between links, which is the permitted maximum value. 

$\eta_{ij,pq}$: normalized cross-correlation between links $ij$ and $pq$ when $l=0$, i.e.,
\begin{equation} \eta_{ij,pq}=\eta_{ij,pq}[0].\end{equation}

 \begin{algorithm}
 \caption{Proposed TRMAC} 
  $\forall a\in V$, node $a$ needs to send data to a neighbor node $b$:
 \begin{algorithmic}[1]
  \STATE \textbf{If} node $a$ has overheard a Pro from node $b$ within $T$, \textbf{go to Step 3}; \textbf{else} $T_{Pro}^{b}=T_{cl}$.
  \STATE Node $a$ transmits the P-R to node $b$ to request for Pro; then, it waits for an interval of no more than $T_{th}$. During the waiting interval, \textbf{if} it receives the Pro from node $b$, \textbf{go to Step 3}; \textbf{else} it retransmits the P-R with a retransmission limit $N_{max}$.
 \STATE \textbf{If} node $a$ has overheard the Pro from other node $j$ ($j \neq b$) within $T_{cl}$, calculate $\eta _{aj, ab}$ based on the received Pro; \textbf{else, go to Step 5}. 
 \STATE \textbf{If} $\left|\eta _{aj, ab}\right|>\eta _{th}$, delay the transmission of the TR-Data by max$\left\{ T_{cl}-T_{Pro}^{j},  T_{cl}-T_{Pro}^{b}, 0\right\}$, \textbf{then go to Step 6}.
 \STATE Delay the transmission of the TR-Data by max$\left\{T_{cl}-T_{Pro}^{b}, 0 \right\}$.
 \STATE Node $a$ transmits the TR-Data to node $b$ by using the time-reversed CIR of the link as a basic waveform.
 \STATE Receiver node $b$ replies the TR-ACK after receiving the TR-Data.
 \STATE \textbf{If} node $a$ has received the TR-ACK within $T_{th}$, \textbf{go to Step 9}.  \textbf{Else}, it retransmits the TR-Data with a retransmission limit $N_{max}$.
 \STATE \textbf{End.}
 \end{algorithmic} 
 \end{algorithm}
  
In TRMAC, $\forall ab\in E$, when node $a$ has data to transmit to node $b$, it accesses the medium in the way described in Algorithm 1. During the handshake of P-R/Pro/TR-Data with Pro/TR-ACK, other nodes that overhear the P-R frame discard it. On the other hand, when overhearing the Pro frame, they save it and record its reception time, in order to take advantage of the Pro in the future. 

As shown in Algorithm 1, when node $a$ needs to transmit data to node $b$,  it first checks whether it has overheard Pro frame from the same node within $T$. If it has received Pro from node $b$ within $T$, node $a$ utilizes the newest received Pro to calculate $g_{ab}[k]$ with omission of the P-R/Pro handshake; then it transmits TR-Data directly when $T_{cl}$ expires, where the backoff interval is max$\left\{T_{cl}-T_{Pro}^{b}, 0 \right\}$, as shown in Step 5 of Algorithm 1. This backoff interval is set to avoid collision with on-going reception at node $b$. As the propagation delay is significantly long in underwater acoustic networks, the omission of handshake of P-R/Pro in this way reduces the end-to-end delay significantly, resulting in an increased channel utilization in MHUANs.

If no previous Pro frame has been overheard from the intended receiver, $T_{Pro}^{b}$ is set to $T_{cl}$ to assure that $T_{cl}-T_{Pro}^{b}=0$ and no backoff is needed for node $b$; node $a$ then conducts the whole handshake of P-R/Pro/TR-Data with Pro/TR-ACK as shown in Fig. 2, by transmitting the P-R to node $b$ for probe requesting. If the channel at receiver node $b$ was already reserved by another link, it defers the reply of Pro frame to node $a$ until the earlier transmission is completed. When node $a$ receives Pro frame from node $b$, it has completed channel reservation and recording, the first step of TRMAC.

During the second step of TRMAC, namely the TR-based link transmission, the removal of collisions caused by the potentially high cross-correlation between links should be ensured. If node $a$ overhears the Pro frame from other node $j  \;(j\neq b)$ within $T_{cl}$, the TR-Data sent by node $a$ may cause collision at node $j$ within $T_{cl}$. It is the heart of the matter whether the transmission from node $a$ to node $b$ will cause collision at node $j$. Equation (15) describes the received signal at node $j$ resulting from transmission of link $ab$, which shows that the induced interference mainly depends on $\eta _{aj, ab}$. 

Thus, we set a threshold for the normalized cross-correlation, $\eta_{th}$, which will be derived in the following subsection. If $\left|\eta _{aj, ab}\right| \leq \eta_{th}$, no collision will be induced at node $j$ by the incoming transmission of link $ab$; node $a$ accesses the channel without considering the induced interference at node $j$, and the backoff interval of max$\left\{T_{cl}-T_{Pro}^{b}, 0 \right\}$ is used. If $\left|\eta _{aj, ab}\right| > \eta_{th}$, node $a$ should defer its transmission of TR-Data frame to avoid the collision at node $j$ and the backoff interval is max$\left\{ T_{cl}-T_{Pro}^{j},  T_{cl}-T_{Pro}^{b}, 0\right\}$, as shown in Step 4 of Algorithm 1. 
    
Due to the significantly improved spatial multiplexing ratio resulting from the active TR process, combined with spatial and temporal uncertainty phenomenon in medium access of MHUANs \cite{ShumingXiong2013RANF,6086556,4509650111}, the proposed TRMAC strategy allows nodes to randomly access medium without carrier sensing, thus leading to an increased channel utilization for MHUANs. By setting a threshold for the normalized cross-correlation between active link and interfering link, the TRMAC effectively removes the residual data collision caused by the high link cross-correlation.

 \subsection{Threshold of Normalized Cross-correlation between Links}
Fig. 1 shows two active links, $ab$ and $ij$, accessing the medium simultaneously. When nodes $a$ and $i$ transmit their TR-Data to nodes $b$ and $j$, respectively, at the same time, with (11) and (14), the total received signal at node $b$ is given by
\begin{equation}\begin{split}
y_{b}[k]&=\left \|\mathbf{h}^{ab}\right \|  x_{a}[k]\quad\quad\quad\;\mbox{(Signal)}\\
&+\left \|\mathbf{h}^{ab}\right \|\sum_{l= 0; \; l\neq (L-1)/D}^{2(L-1)/D}\eta_{ab,ab}[Dl-(L-1)] x_{a}[k-l] \mbox{(ISI)}\\
&+\left \|\mathbf{h}^{ib}\right \|\sum_{l=0}^{2(L-1)/D}\eta_{ib,ij}[Dl-(L-1)] x_{i}[k-l] \quad\mbox{(ILI)}\\
&+n^{'}[Dk],  \quad\quad\quad\quad\quad\quad\; \mbox{(Noise)}
\end{split}\end{equation}
where $\left\{n^{'}[Dk]\right\}$ is the sequence of the complex-valued additive white Gaussian noise samples with zero-mean and variance $\sigma^{2}$,  and ILI is caused by the simultaneous transmissions. Then one can get the received ILI power as
\begin{equation}\begin{split}
P_{\texttt{ILI}}&= DP\left \|\mathbf{h}^{ib}\right \|^{2}\left|\eta_{ib,ij}\right|^2\\ 
&+DP\left \|\mathbf{h}^{ib}\right \|^{2}\sum_{l= 0; \; l\neq (L-1)/D}^{2(L-1)/D}\left|\eta_{ib,ij}[Dl-(L-1)]\right|^{2},\end{split}\end{equation}
where $0 \leq \left|\eta_{ib,ij}\right|<1$. It is assumed that the average transmit power is the same at each node in the network.

The effective SINR of the active TR-based simultaneous transmissions (ATRSTs) in TRMAC is defined as 
\begin{equation}
\gamma^{\texttt{ATRSTs}}=\frac{P_{\texttt{Sig}}}{P_{\texttt{ISI}}+P_{\texttt{ILI}}+\sigma^{2}}.\end{equation}
As shown in (13) and (17), with the increase of factor $D$, the sum of $P_{\texttt{ISI}}$ and $P_{\texttt{ILI}}$ is finally reduced to $DP\left \|\mathbf{h}^{ib}\right \|^{2}\left|\eta_{ib,ij}\right|^2$. In Section V, the impact of the factor $D$ on the effective SINR will be evaluated over the acoustic channels modeled by BELLHOP\cite{bellhop}, which can provide detailed and realistic modeling of the underwater acoustic channel through the measured marine environment parameters\cite{oceans17zhao}.

With a minimum required SINR at the receiver, i.e. $\gamma^{\texttt{ATRSTs}}\geq \gamma$, and using (16) and (18), one can obtain (19),
\begin{figure*}[h]\begin{equation} \left|\eta_{ib,ij}\right| \leq 
\left(\frac{\left \|\mathbf{h}^{ab}\right \|^{2}/\gamma-\left \|\mathbf{h}^{ab}\right \|^{2}\sum\limits_{l=0;\; l\neq(L-1)/D}^{2(L-1)/D}\left|\eta_{ab,ab}[Dl-(L-1)]\right|^{2}-\left \|\mathbf{h}^{ib}\right \|^{2}\sum\limits_{l=0;\; l\neq(L-1)/D}^{2(L-1)/D}\left|\eta_{ib,ij}[Dl-(L-1)]\right|^{2}-\sigma^{2}/DP}{\left \|\mathbf{h}^{ib}\right \|^{2}}\right)^{1/2},\end {equation} \end{figure*}
which shows the threshold of the normalized cross-correlation between the interfering link and other active link. When the transmission from node $a$ to node $b$ is ahead of the one from node $i$ to node $j$, node $i$ should check whether (19) is satisfied. If it is, the incoming transmission from node $i$ to node $j$ will not interfere with the reception at node $b$. During the calculation of this threshold, node $i$ obtains the CIRs of links $ib$ and $ij$ through the reception of the Pro frames from nodes $b$ and $j$, respectively. Since node $b$ can get the CIR of link $ab$ during the channel reservation and recording stage, $\left \|\mathbf{h}^{ab}\right \|$ and $\eta_{ab,ab}[Dl-(L-1)]$ required in (19) can be calculated and then piggybacked on the Pro frame, which enables node $i$ to calculate the threshold of the normalized cross-correlation. 

Similarly, if the transmission on link $ij$ is ahead of that on link $ab$, as described in Algorithm 1, node $a$ exploits its threshold of the normalized cross-correlation, i.e. $\eta_{th}$, and checks the value of $\left|\eta_{aj,ab}\right|$. If $\left|\eta_{aj,ab}\right|$ exceeds this threshold, its transmission will interfere with the on-going transmission on the link $ij$. In this case, node $a$ defers its transmission of TR-Data. Otherwise, it means that the cross-correlation between the two links is low enough to ensure their simultaneous transmissions, and node $a$ transmits TR-Data to node $b$ directly. 

As the threshold of the link cross-correlation is set under the assumption that the received SINR is not lower than a minimum required SINR $\gamma$ at the receiver, no collision is caused by the near-far problem in the TRMAC. With the assumption of the same transmit power for each node, in Fig. 1, the link $ij$ may induce severe interference to the link $ab$ at node $b$, when node $i$ is much nearer to $b$ than node $a$. This near-far problem is caused by the fact that $\left \|\mathbf{h}^{ib}\right \|^{2} $ is much larger than $\left \|\mathbf{h}^{ab}\right \|^{2}$, which results in a significant low value of the threshold of the normalized cross-correlation, as can be seen from (19).
Through the strategy of the cross-correlation threshold, the simultaneous transmissions of the links $ij$ and $ab$ would be prevented by deferring one link to resolve the near-far problem.

The strategy of the cross-correlation threshold effectively prevents potential collisions induced by the near-far problem and the high cross-correlation between links in the proposed TRMAC protocol. In this way, the spatial and temporal focusing, which is the nature of the active TR process, is exploited by the proposed TRMAC strategy to improve the spatial multiplexing ratio and channel utilization in MHUANs with an improved successful transmission ratio at the data link layer.  

\section{Performance Evaluation}
In this section, the performance of the proposed TRMAC protocol is evaluated through extensive simulations. Due to the complex characteristics of the underwater acoustic channel, the underwater acoustic network environment is simulated by combining the OPNET Modeler and BELLHOP\cite{bellhop}. BELLHOP provides detailed and realistic modeling of the underwater acoustic channel through the measured marine environment parameters. In this section, a sound velocity profile measured during the sea trial at South China Sea in September 2014, as shown in Fig. 3, is used in the BELLHOP simulations to simulate acoustic ray trajectories and propagation delays. 
\begin{figure}[h] 
\centering
\includegraphics[width=3.4 in]{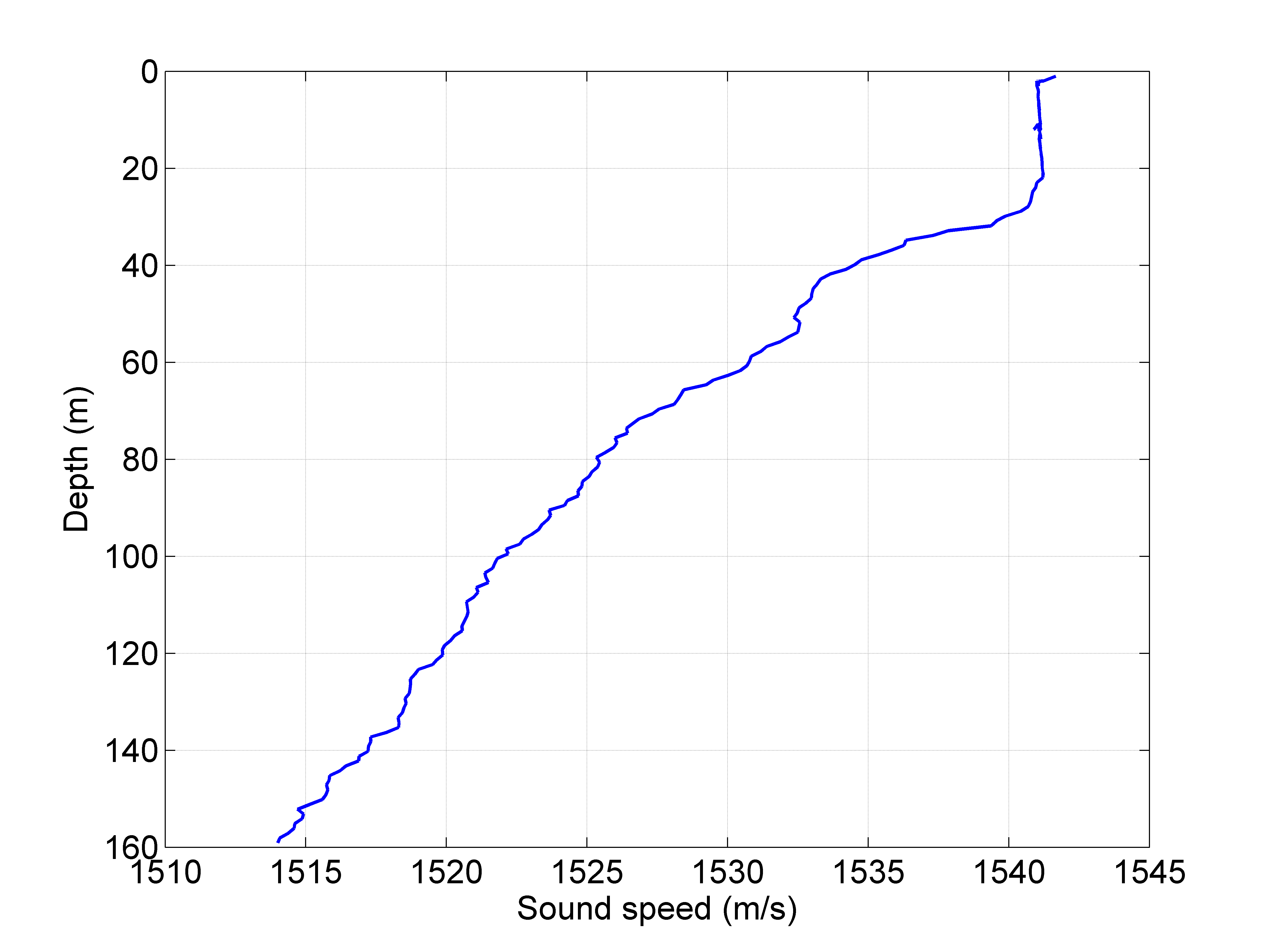}
\caption{Sound velocity profile.}
\label{fig3}
\end{figure}
\subsection{Effective SINR with Factor $D$}
Firstly, the ATRSTs is compared with the single direct transmission (SDT) without TR. As shown in Fig. 1, both links $ab$ and $ij$ have data packets to transmit. In the SDT, without the interference isolation achieved by the active TR process, only one link can be active at a time in the neighborhood. Without using the TR-based transmission at the physical layer, the received signal of the SDT over link $ab$ can be expressed as
\begin{equation}\begin{split}
y^{\texttt{SDT}}[k]
&=\sum_{l=0}^{(L-1)/D}h_{ab}[Dl] x_{a}[k-l]+n[Dk]\\
&=h_{\bar{l}}^{ab} x_{a}[k-\bar{l}/D] \quad\quad\quad\quad\mbox{(Signal)}\\
&+\sum_{l=0; l\neq \bar{l}/D}^{(L-1)/D}h_{ab}[Dl] x_{a}[k-l] \quad\mbox{(ISI)}\\
&+n[Dk],\quad\quad\quad\quad\quad\quad\quad\quad\mbox{(Noise)}
\end{split}\end{equation} 
where $h_{\bar{l}}^{ab}$ is the strongest tap of the CIR of link $ab$, and $\bar{l}$ is assumed to be dividable by $D$ for simplicity. Due to the multi-path effect, the up/down-sampling method is also used in the SDT transmission to reduce the ISI in (20). As no interfering link appears in the SDT transmission, the effective SINR of the SDT over link $ab$ can be expressed as 
\begin{equation} \gamma^{\texttt{SDT}}=\frac{DP \left|h_{\bar{l}}^{ab}\right|^{2}}{DP \sum\limits_{l=0; \;l\neq \bar{l}/D}^{(L-1)/D}\left|h_{ab}[Dl]\right|^{2}+\sigma^{2}}. \end {equation}

Next, the impact of the factor $D$ on the effective SINR is studied based on the CIRs of underwater acoustic links modeled using BELLHOP with $L$ = 130 and carrier frequency $f$ = 25 kHz. The ATRSTs is evaluated in Figs. 4 and 5, where the network topology shown in Fig. 1 is used with the positions (depth(m), distance (m)) of the nodes set as: $a$ (20, 0), $b$ (20, 1000), $i$ (50, 0), and $j$ (70, 1000). Note that $P_{a}$ is the acoustic power translated from the average electric transmit power $P$, and $\sigma_{a}^{2}$ is the acoustic power of the ambient noise.

Fig. 4 shows the effective SINR of link $ab$ in the ATRSTs and the SDT, respectively, where both links $ab$ and $ij$ are active in the ATRSTs and only one link $ab$ is working in the SDT. It demonstrates that the ATRSTs outperforms the SDT under low $P_{a}/\sigma_{a}^{2}$, which shows the effectiveness of the TRMAC. Due to the interference reduction caused by the up/down-sampling method, the effective SINRs in both ATRSTs and SDT increase with the factor $D$, as shown in Fig. 4. Note that although a large $D$ ($D > 1$) diminishes ISI and ILI significantly, it requires $D$ times the bandwidth when compared with $D$ = 1, under the assumption of the same data rate. Thus, there is a tradeoff between interference reduction and bandwidth utilization. On the other hand, due to the spatial focusing of the active TR, the ISI and ILI power is lower than the signal power in the ATRSTs, which results in a reasonable effective SINR with a moderate value of $D$.

Fig. 5 presents the impact of $\left|\eta_{ib,ij}\right|$ on the effective SINR in the ATRSTs with $P_{a}/\sigma_{a}^{2}$ = 65 dB, and shows that the effective SINR at node $b$ decreases with $\left|\eta_{ib,ij}\right|$. According to (15) and (17), the impact of $\left|\eta_{ib,ij}\right|$ on the ILI increases with $D$, resulting in an increased dependance of the effective SINR on $\left|\eta_{ib,ij}\right|$ with $D$. Therefore, the threshold design in the TRMAC for the normalized cross-correlation between links can resolve the potential collisions caused by the potentially high link cross-correlation.

\begin{figure}[h] 
\centering
\includegraphics[width=3.4 in]{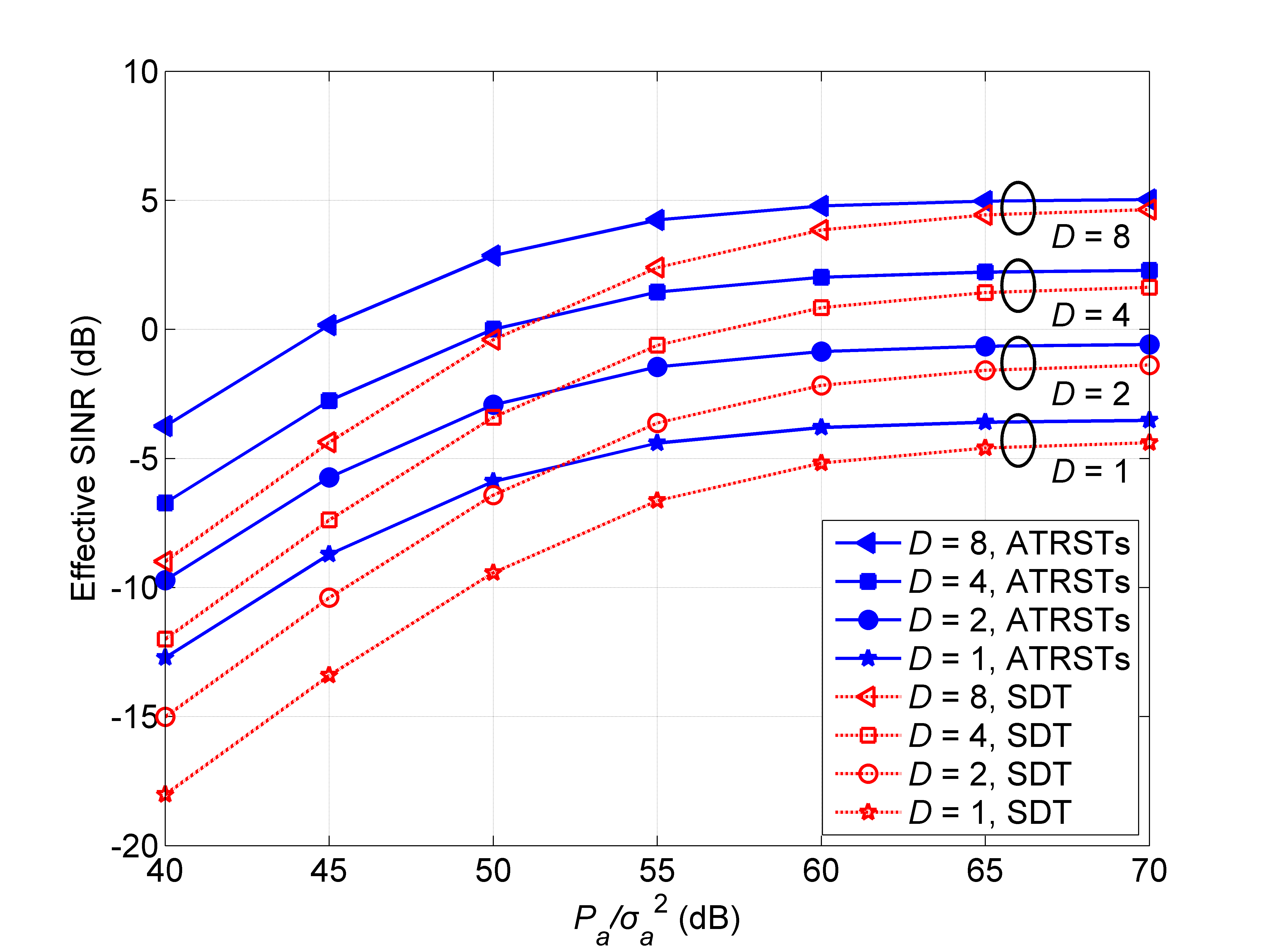}
\caption{ATRSTs vs. SDT (solid line: ATRSTs; dashed line: SDT)}
\label{fig16}
\end{figure}
\begin{figure}[htbp] 
\centering
\includegraphics[width=3.4 in]{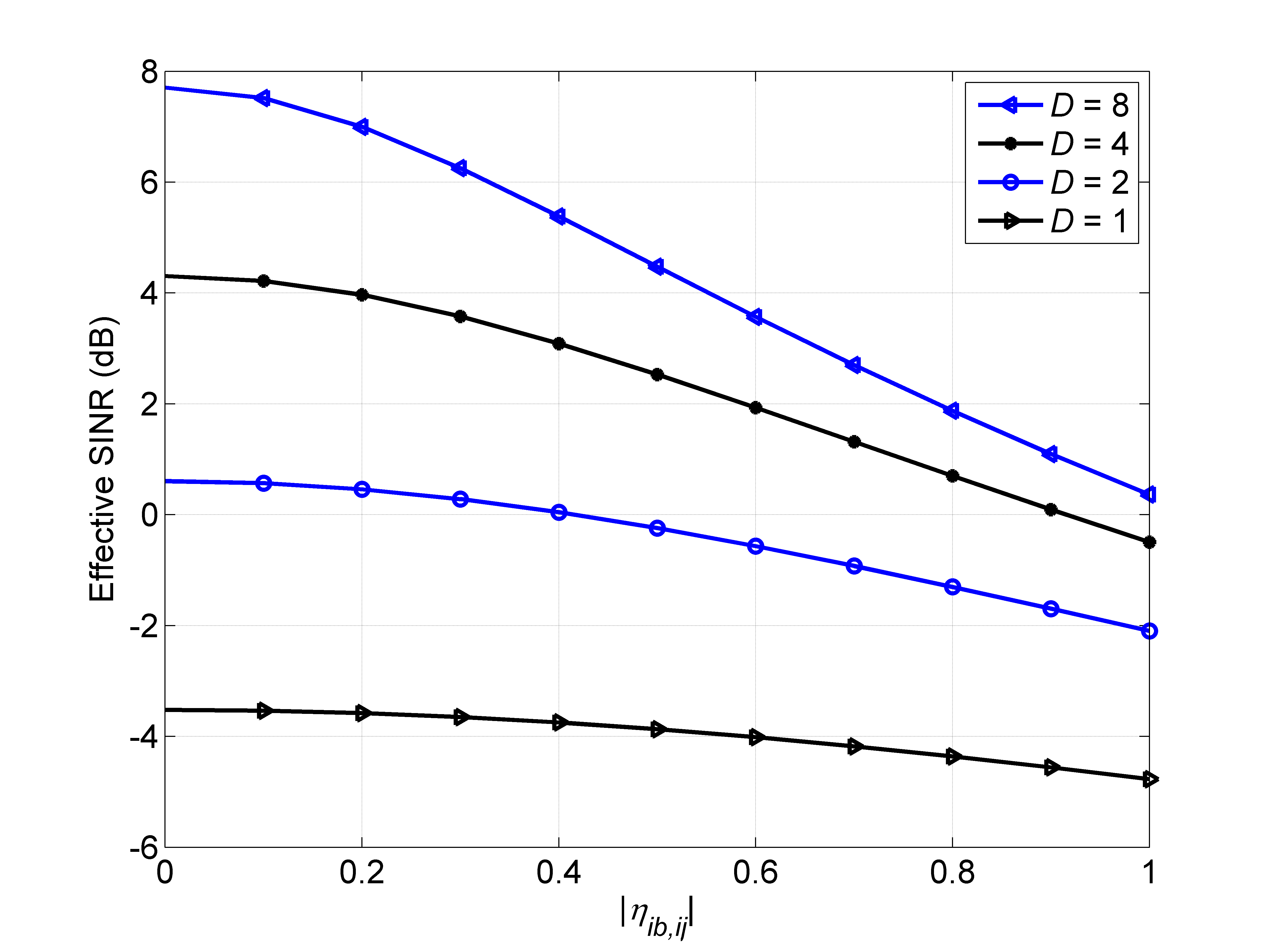}
\caption{The impact of $\left|\eta_{ib,ij}\right|$ on the ATRSTs when $P_{a}/\sigma_{a}^{2}$ = 65 dB.}
\label{fig15}
\end{figure}

\subsection{Normalized Cross-correlation between Links}
To illustrate the link cross-correlation of MHUANs, $\left|\eta_{ip,ij}\right|$ is obtained based on (4), (15), and the CIRs are modeled with BELLOP. With the positions of the reference transmitter $i$ and receiver $j$ fixed, $\left|\eta_{ip,ij}\right|$ is calculated by changing the position of node $p$, as shown in Figs. 6 and 7. Note that in Fig. 6, the positions of nodes $i$ and $j$ are the same as in the above subsection; in Fig. 7, their positions are (100,0) and (100, 4000), respectively; and in these figures, the positions of nodes $i$ and $j$ are indicated by red and yellow nodes, respectively.
\begin{figure}[htbp] 
\centering
\includegraphics[width=3.2 in]{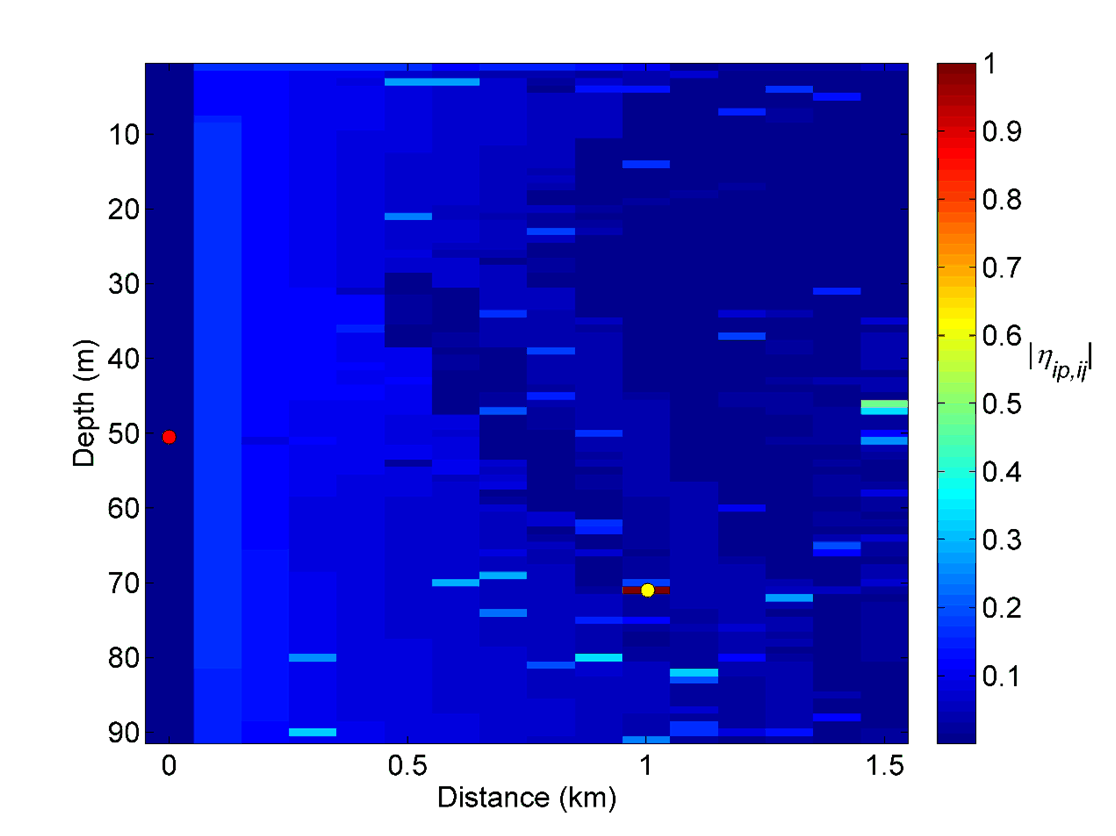}
\caption{$\left|\eta_{ip,ij}\right|$ (reference link: $ij$; positions (depth(m), distance (m)): $i$ (50, 0), $j$ (70, 1000)).}
\label{fig4}
\end{figure}
\begin{figure}[htbp] 
\centering
\includegraphics[width=3.2 in]{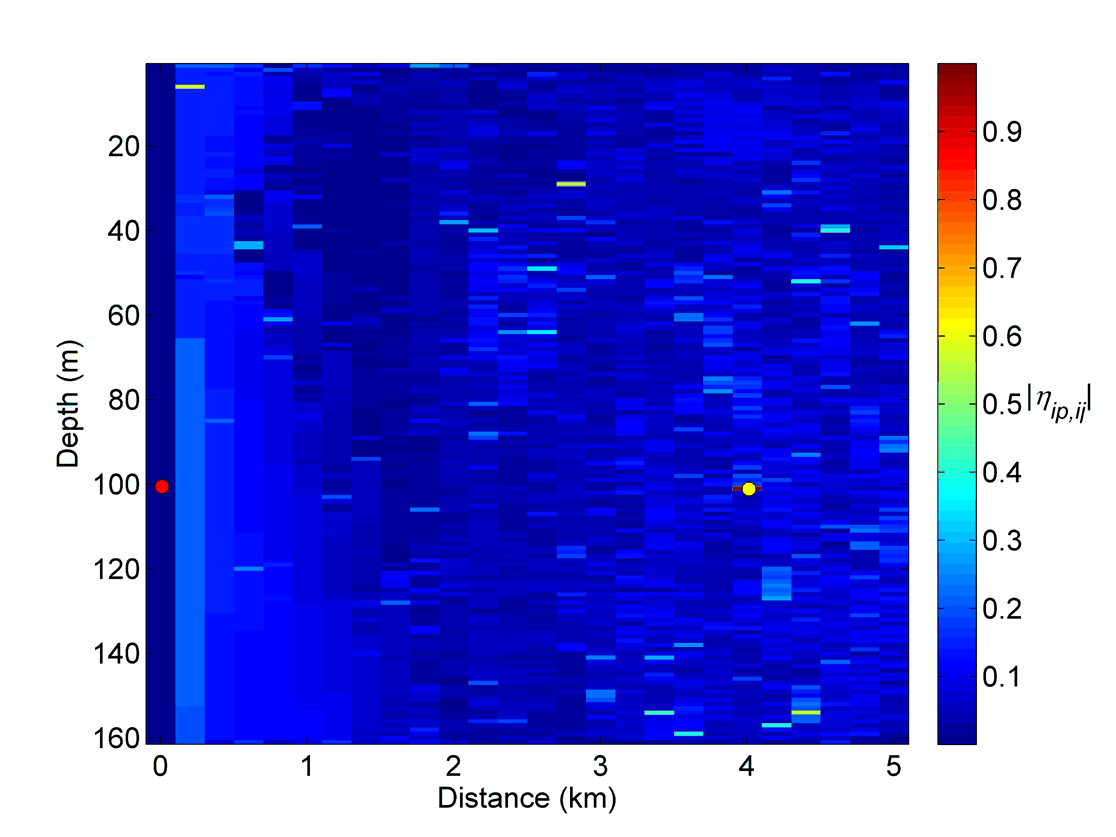}
\caption{$\left|\eta_{ip,ij}\right|$ (reference link: $ij$; positions (depth(m), distance (m)): $i$ (100, 0), $j$ (100, 4000)).}
\label{fig5}
\end{figure}
From Figs. 6 and 7, one can see that the cross-correlation between links over the underwater region is low in most cases, which confirms the spatial variability of the underwater acoustic links. The high cross-correlation potentially appears only when the link length and the depths of the transmitter /receiver are similar to those corresponding to the reference link $ij$.

Typically, to cover a broad underwater region, MHUAN nodes are deployed sparsely at diverse depths and spaced at various distances, leading to low cross-correlation between different links. This feature is exploited by TRMAC to isolate the interference of simultaneous active TR-based transmissions. Furthermore, by calculating the normalized cross-correlation of links and setting a threshold for it, TRMAC can remove the residual collision caused by a potentially high cross-correlation. 

\subsection{MAC Protocol Performance}
We consider a multi-hop network topology, where 20 nodes are randomly distributed in a 4 km $\times$ 4 km region with node depth ranging from 0 m to 50 m, constituting an MHUAN. These nodes can form a maximum of 10 active links accessing the channel simultaneously in the multi-hop network. Packet generation at each link is subject to the exponential distribution with the mean packet arrival interval of 8 s. Due to the complex characteristics of the underwater acoustic channel, the underwater acoustic network environment is simulated by combining the OPNET Modeler and BELLHOP. The simulation parameters are given in Table 1.

\begin{table}[h]
\caption{Simulation Parameters.}
\label{table}
\centering
\setlength{\tabcolsep}{3pt}
\begin{tabular}{|c|c|}
\hline
Parameter& 
Value \\
\hline
Depth of water & 80 m\\
Network deployment region size  & 4 km $\times$ 4 km\\
Node depth & 0 - 50 m\\
1-hop range  & 1 km\\
Node number & 20\\
Maximum number of simultaneous active links & 10\\
Maximum number of hops & 6\\
Data rate & 512 bps\\
$D$ & 4\\
Bandwidth  & 4 kHz\\
Mean of packet arrival interval & 8 s\\
Packet length & 256 bits\\
Constant maximum back-off interval for S-CSMA/CA & 2 s\\
$T$ & 30 s\\
$\Delta$ & 0.25 s\\
$N_{max}$ & 3\\
\hline
\end{tabular}
\label{tab1}
\end{table}

As the proposed TRMAC is a random access protocol, we compare it with the conventional CSMA/CA protocol \cite{820738}. Due to the large propagation delay in MHUAN, the spatial and temporal uncertainty problem worsen the network performance significantly. To study the impact of this problem on the traditional CSMA/CA, its binary exponential back-off algorithm is replaced by setting a constant maximum back-off interval; this modified CSMA/CA is referred to as the simplified CSMA/CA (S-CSMA/CA).

The following three performance metrics are used to evaluate the TRMAC protocol:
\begin{itemize}
\item Delay: the interval from the time the data packet was generated by the transmitter node at application layer to the time it is received correctly;
\item Data drop ratio: the ratio of the number of dropped data frames to that of data frames that have been transmitted, which is an indicator of the collision ratio;
\item Throughput: the rate of receiving data correctly at the receiver node when the link is busy. 
\end{itemize}

In order to evaluate the performance of the MAC protocols under different network loads, 10, 8, 6, and 4 active links are chosen respectively during simulations. With 10 links, all 20 nodes of the network are engaged and the network load is the highest. Figs. 8-10 show the delay, data drop ratio, and throughput over the simulation time when the proposed TRMAC protocol is adopted in MHUANs. When 10 active links are chosen in the network, although the data drop ratio increases with the network load, it is still low (average value is less than 0.08)  because of the collision minimization mechanism in the proposed TRMAC protocol.

Similarly, Figs. 11-13 present the network performance under the heavy network load of the TRMAC, CSMA/CA and S-CSMA/CA protocols with 10 active links, which show that the TRAMC performs better than the CSMA/CA and S-CSMA/CA protocols. CSMA/CA does not perform well in underwater acoustic networks due to the ineffective carrier sensing under the circumstance of a large propagation delay. However, by exploiting the interference isolation resulting from the active TR process, the proposed TRMAC strategy allows nodes to randomly access medium without carrier sensing, thus leading to an increased channel utilization for MHUANs. 

As shown in Fig. 11, under the heavy network load, the data drop ratio is very high when the CSMA/CA mechanisms are used at the MAC layer. Additionally, the delay increases quickly with the simulation time and the throughput is limited in both CSMA/CA mechanisms. Due to the heavy network load setting in Figs. 11-13, both CSMA/CA and S-CSMA/CA protocols are faced with high collision ratio, resulting in a large number of retransmissions among the 10 active links. Thus, the new arriving packets at the 10 transmitters will face long queuing delay. The queue of each transmitter will grow with the simulation time, resulting in a growing delay for each packet with the simulation time, as shown in Fig.12. Consequently, as shown in Fig. 13, the CSMA/CA mechanisms have low network throughput under the 10 active links.

Figs. 14-16 show the performance comparison of the three MAC protocols with 10, 8, 6, and 4 active links, respectively. As shown in these figures, the proposed TRMAC outperforms CSMA/CA and S-CSMA/CA under different network loads. This is mainly due to the improved channel utilization resulting from the active TR process at the physical layer. The cross-layer design of the TRMAC strategy allows different active links to access the channel at the same time with minimized collision. Therefore, the network delay and packet drop ratio are reduced greatly in TRMAC when compared with CSMA/CA mechanisms. Further, the throughput is improved significantly when the TRMAC is utilized in the network. From the simulation results of the two CSMA/CA mechanisms, one can see that S-CSMA/CA does not perform better than the traditional CSMA/CA. Note that the maximum back-off interval is $2^{i}$ s for the $i$-th retransmission in CSMA/CA, while its value is 2 s in S-CSMA/CA. By setting a smaller constant maximum back-off interval, S-CSMA/CA may reduce the delay and improve the throughput; however, this results in increased data frame collisions especially under higher network load.

Though both TRMAC and CSMA/CA mechanisms are random access protocols based on the handshake of four frames, they perform differently because of their corresponding techniques at the physical layer and the novel cross-layer design for TRMAC. Simulation results shown in this section have verified the capability of the active TR to exploit the multi-path energy from the richly scattering underwater environment and isolate the interference between adjacent links.

\begin{figure}[htbp] 
\centering
\includegraphics[width=3.5 in]{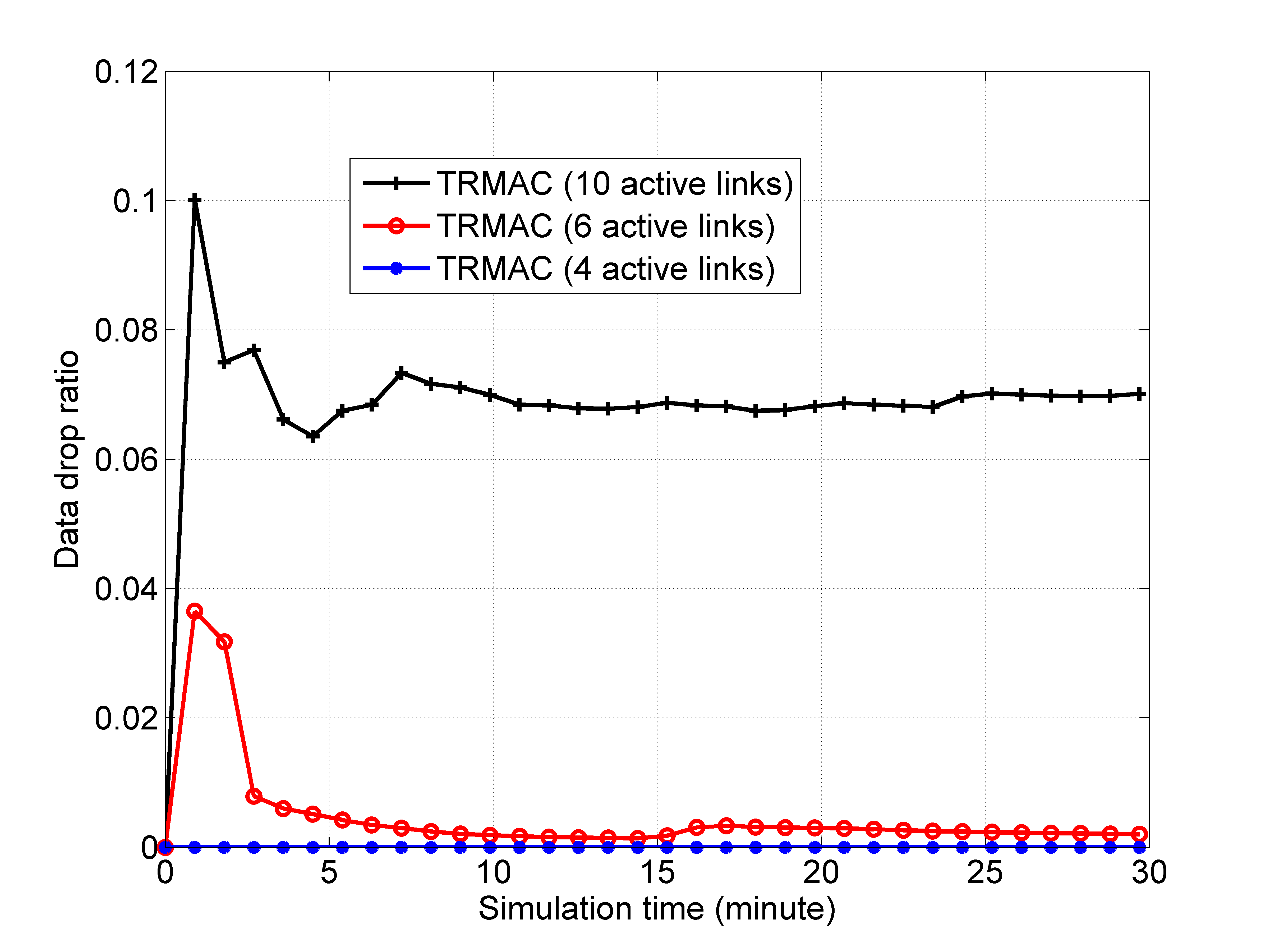}
\caption{Data drop ratio of TRMAC for different numbers of active links.}
\label{fig6}
\end{figure}
\begin{figure}[htbp] 
\centering
\includegraphics[width=3.5 in]{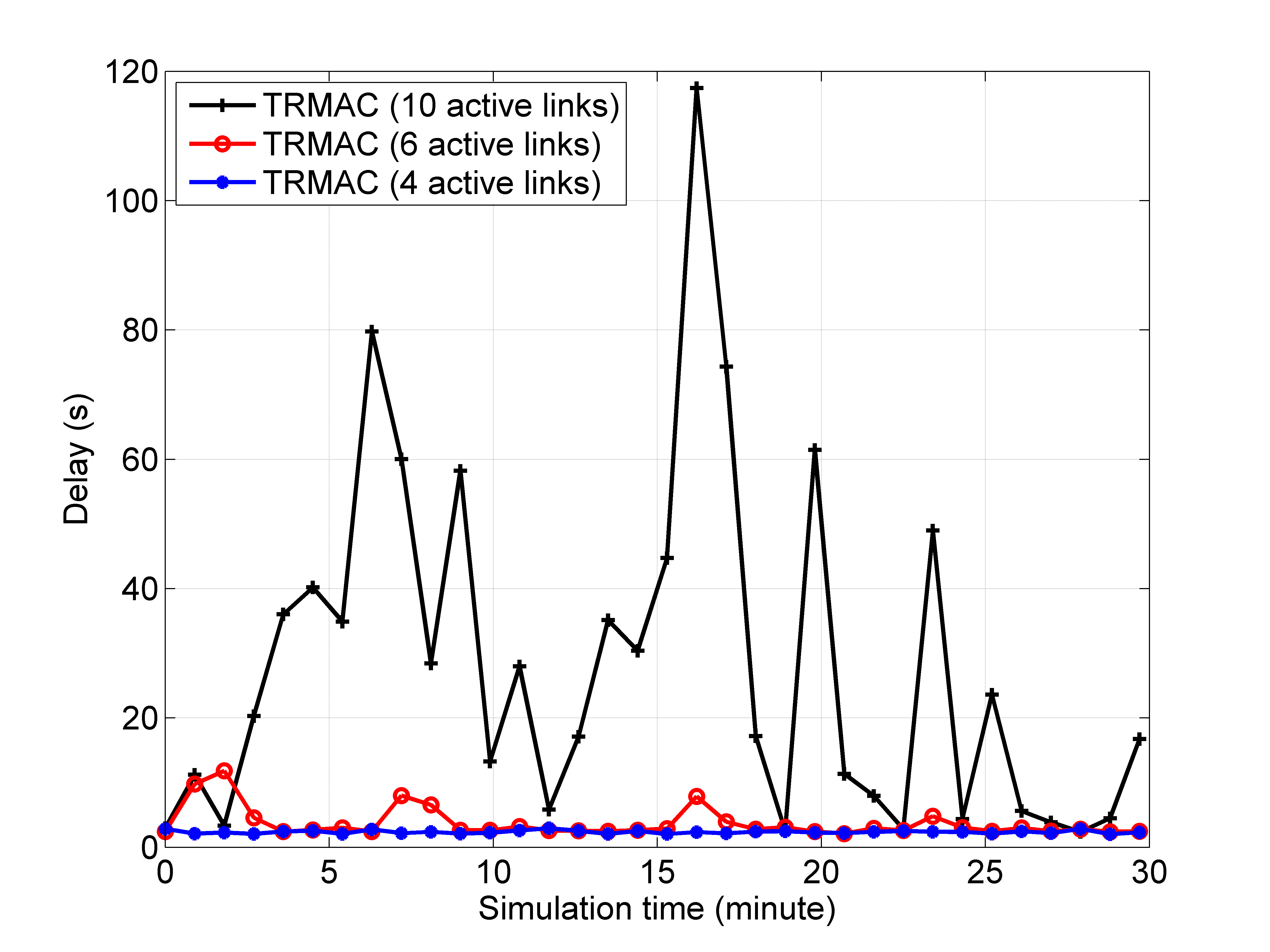}
\caption{Delay of TRMAC for different numbers of active links.}
\label{fig7}
\end{figure}
\begin{figure}[htbp] 
\centering
\includegraphics[width=3.5 in]{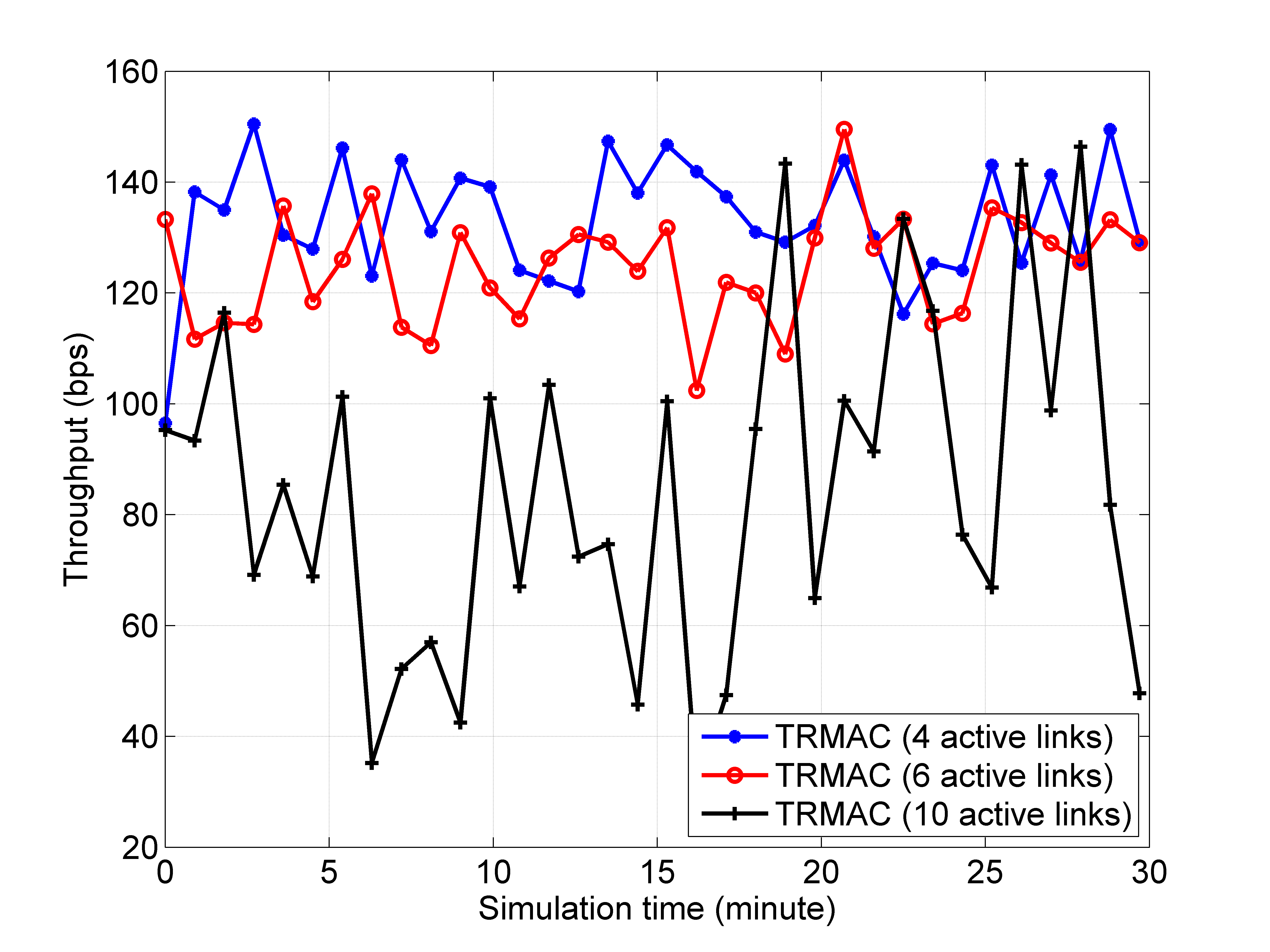}
\caption{Throughput of TRMAC for different numbers of active links.}
\label{fig8}
\end{figure}
\begin{figure}[htbp] 
\centering
\includegraphics[width=3.5 in]{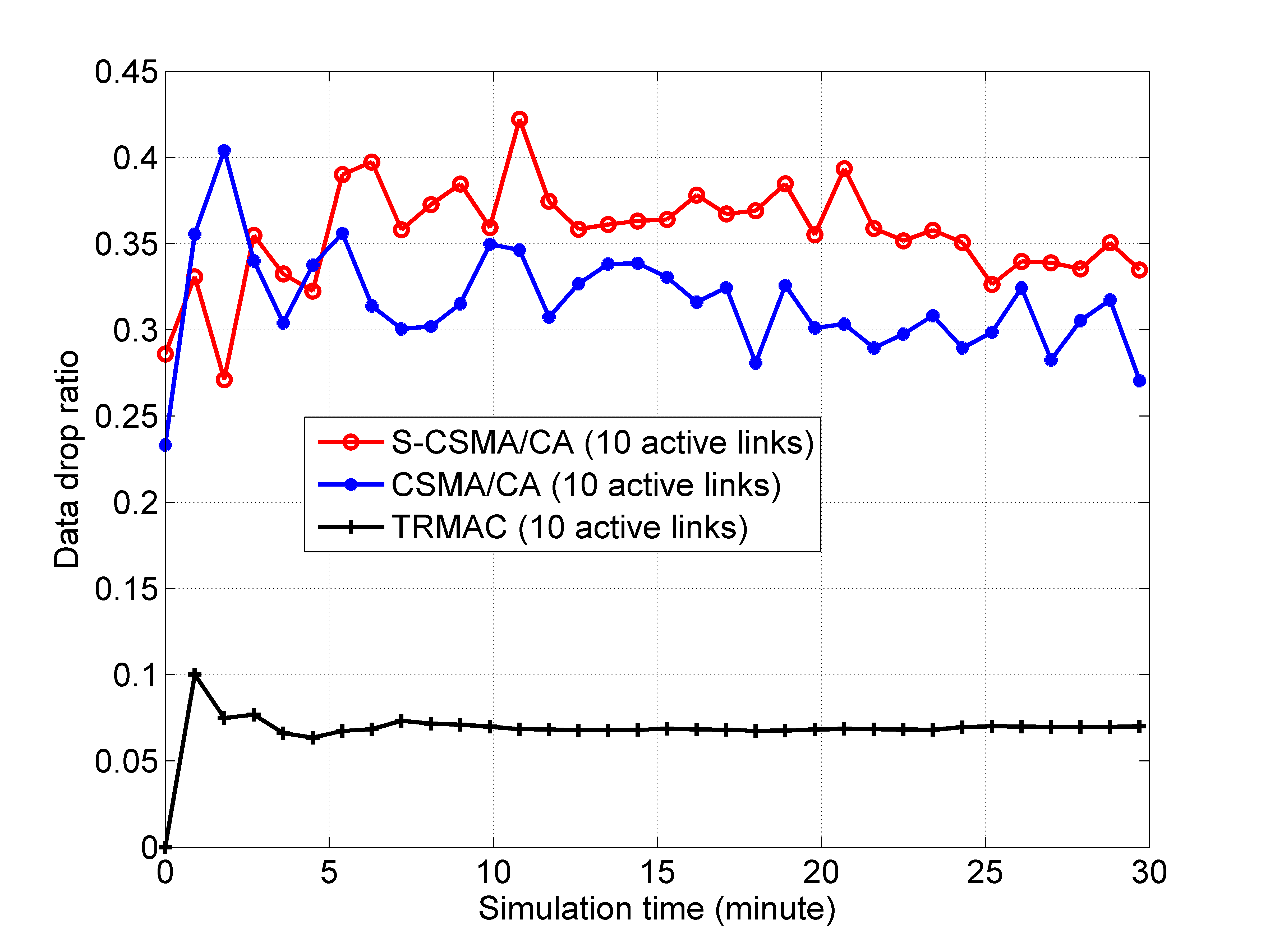}
\caption{Data drop ratio of TRMAC, CSMA/CA and S-CSMA/CA for 10 active links.}
\label{fig9}
\end{figure}
\begin{figure}[htbp] 
\centering
\includegraphics[width=3.5 in]{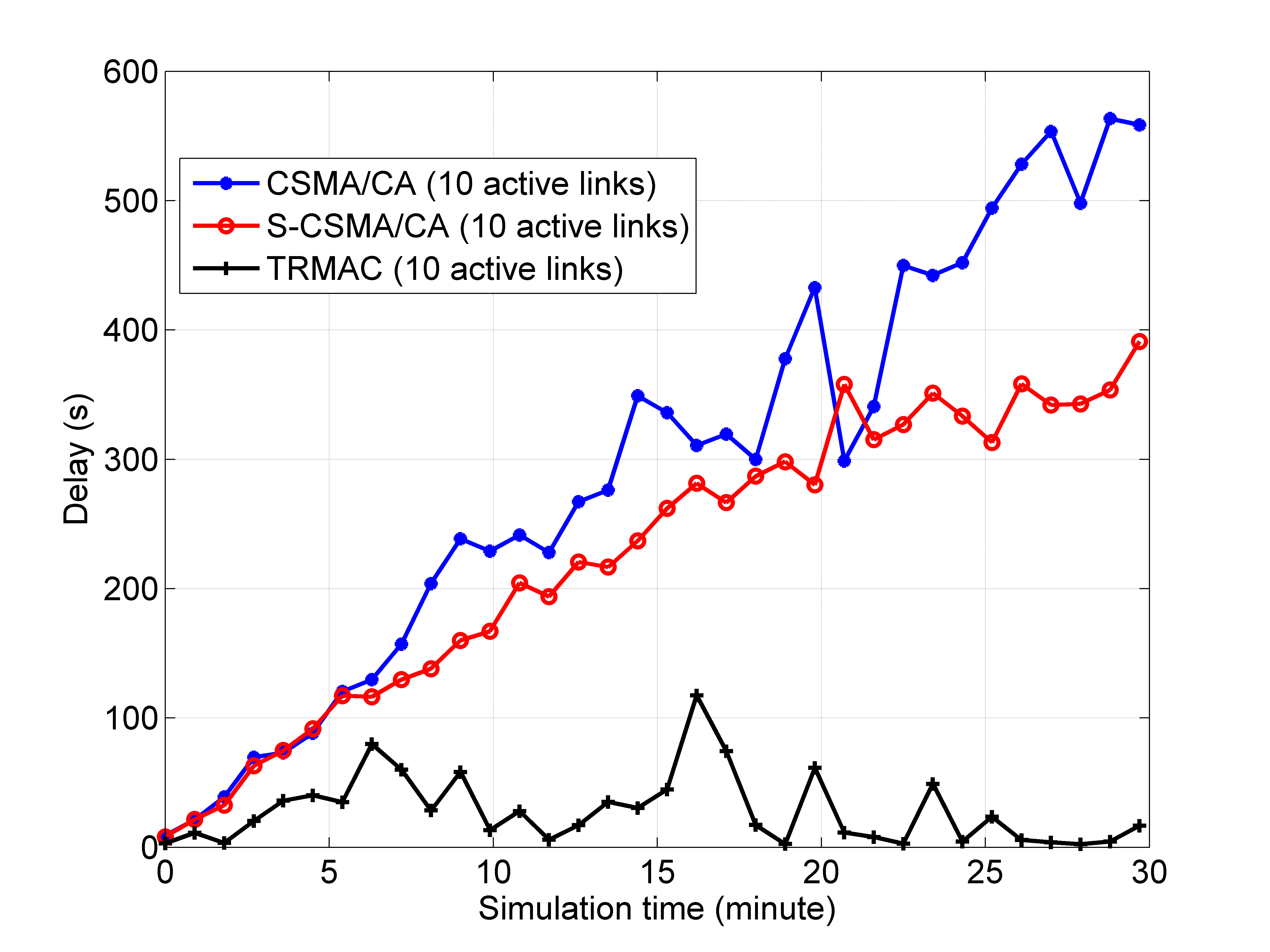}
\caption{Delay of TRMAC, CSMA/CA and S-CSMA/CA for 10 active links.}
\label{fig10}
\end{figure}
\begin{figure}[htbp] 
\centering
\includegraphics[width=3.5 in]{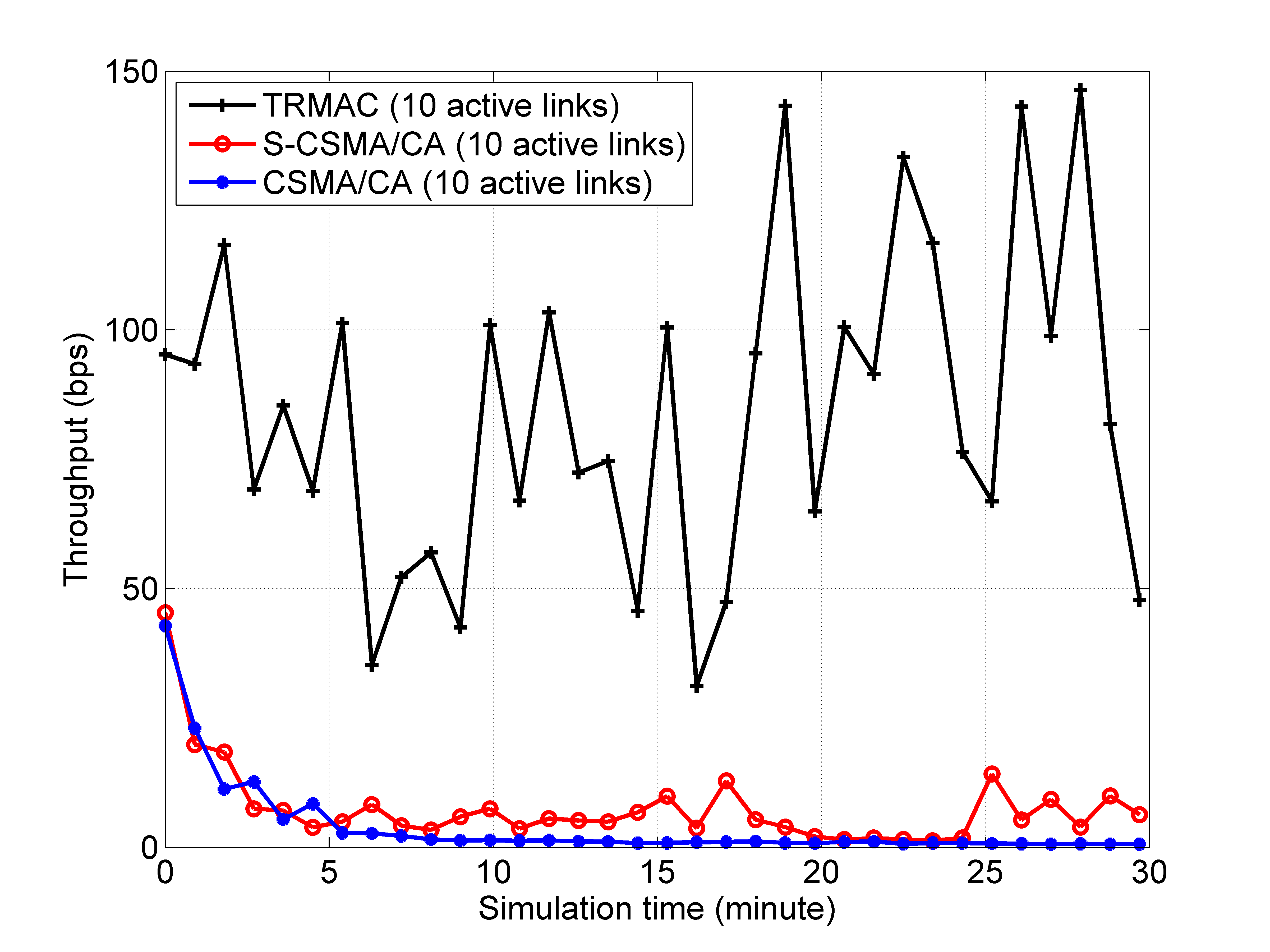}
\caption{Throughput of TRMAC, CSMA/CA and S-CSMA/CA for 10 active links.}
\label{fig11}
\end{figure}
\begin{figure}[htbp] 
\centering
\includegraphics[width=3.5 in]{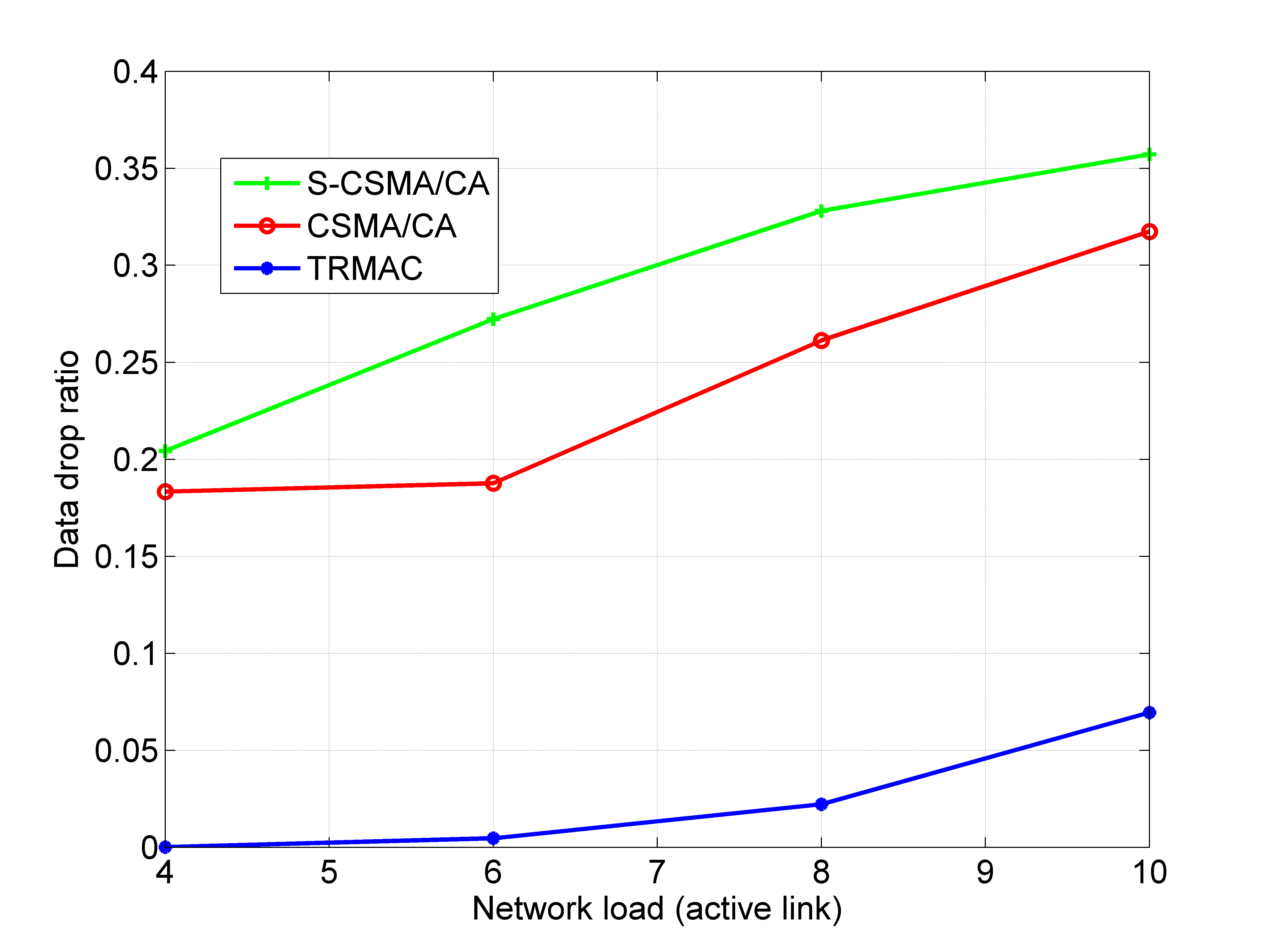}
\caption{Data drop ratio of TRMAC, CSMA-CA and S-CSMA/CA  versus the network load.}
\label{fig12}
\end{figure}
\begin{figure}[htbp] 
\centering
\includegraphics[width=3.5 in]{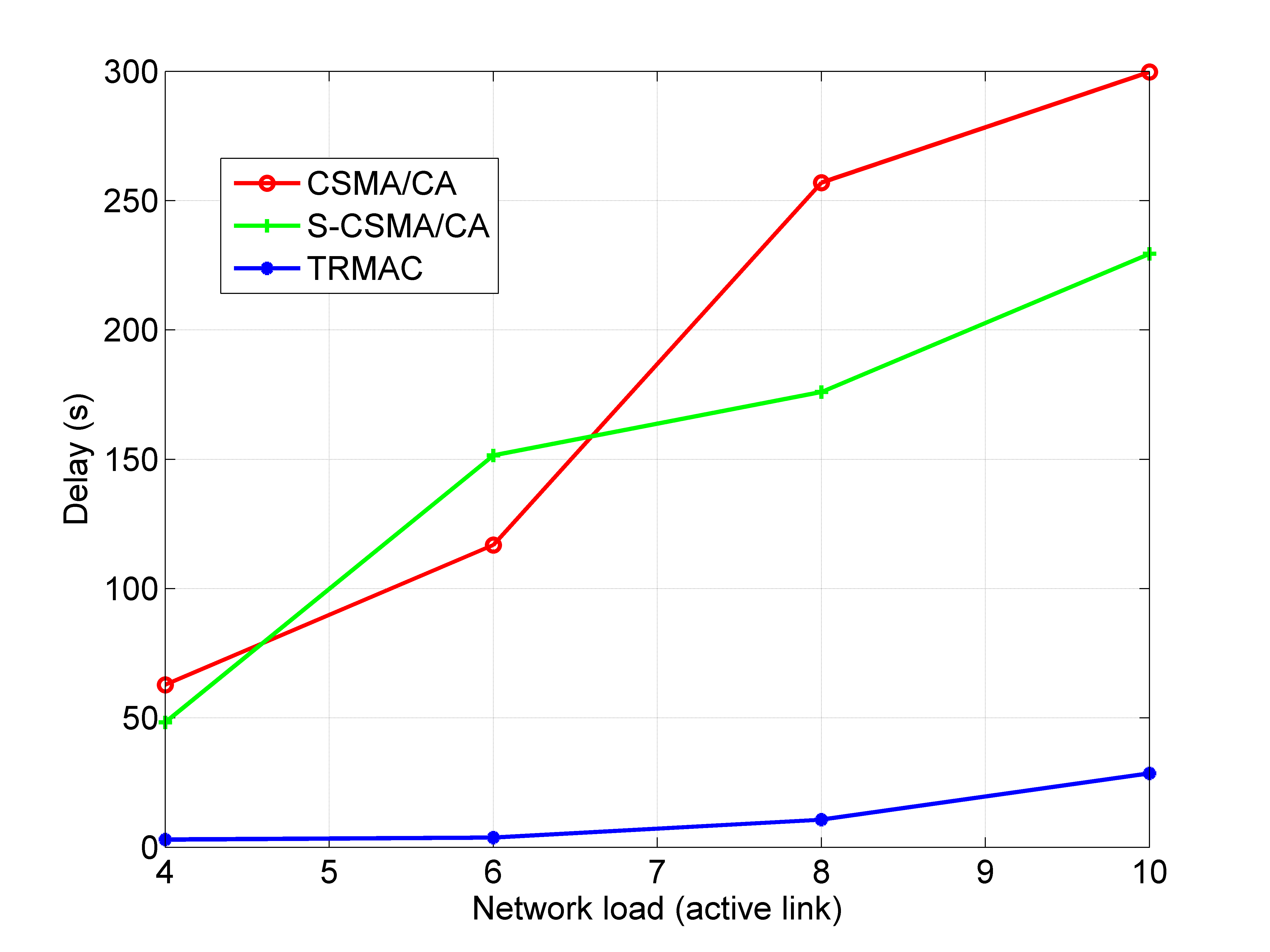}
\caption{Delay of TRMAC, CSMA-CA and S-CSMA/CA versus the network load.}
\label{fig13}
\end{figure}
\begin{figure}[htbp] 
\centering
\includegraphics[width=3.5 in]{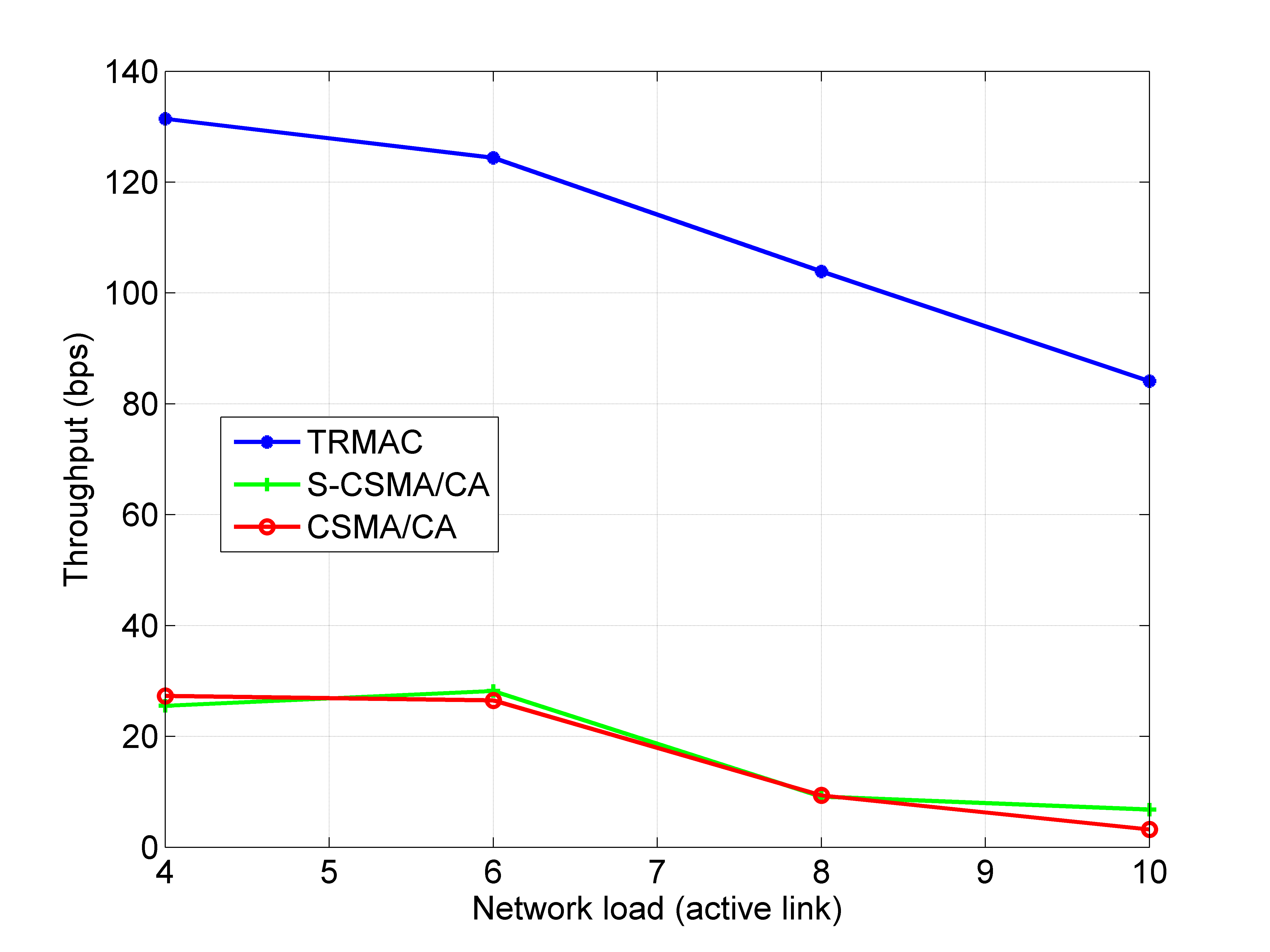}
\caption{Throughput of TRMAC, CSMA-CA and S-CSMA/CA versus the network load.}
\label{fig14}
\end{figure}

\section{Conclusions}
Due to its capability of exploiting the multi-path energy from the richly scattering underwater environment and making the signal energy focus in both spatial and temporal domains, the active time reversal (TR) process plays an important role in the medium access control (MAC) mechanism. We showed that the active TR process can isolate interference between adjacent links and avoid transmission collision in MHUANs. An active TR-based MAC (TRMAC) protocol is proposed for MHUANs with the aim of minimizing collision and maximizing channel utilization simultaneously. Furthermore, considering the impact of cross-correlation between different links on medium access, the threshold of the link cross-correlation is derived to resolve the collision caused by the potentially high cross-correlation between links with similar distance and depth. Simulation results show significant increased throughput with decreased delay and data drop ratio when the TRMAC is used, which verifies the effectiveness of the proposed protocol. 

As the TR process is suitable for underwater acoustic networks, it can be utilized in different styles at the MAC layer. For instance, it can be combined with TDMA, CDMA, or other random MAC protocols. This will be investigated in future work in order to exploit the inherent nature of the active TR process in underwater acoustic networks.

\bibliographystyle{IEEEtran}

\bibliography{ref}

\ifCLASSOPTIONcaptionsoff
  \newpage
\fi

\end{document}